\documentclass[a4paper,11pt]{article}
\pdfoutput=1 

\usepackage{jcappub} 
\usepackage{pgfplots}
\usepackage{tikz}
\usetikzlibrary{matrix}
\usepackage[T1]{fontenc} 
\usepackage[autostyle]{csquotes}

\title{\boldmath Are there really conformal frames?\\ Uniqueness of affine inflation}

\author{Hemza Azri}

\affiliation[]{Department of Physics, \.{I}zmir Institute of Technology, \\ TR35430
\ \.{I}zmir, Turkey}

\emailAdd{hemzaazri@iyte.edu.tr; hm.azri@gmail.com}

\abstract{Here we concisely review the nonminimal coupling dynamics of a single scalar field in the context of purely affine gravity and extend the study to multifield dynamics. The coupling is performed via an affine connection and its associated curvature without referring to any metric tensor. The latter arises \textit{a posteriori} and it may gain an emergent character like the scale of gravity. What is remarkable in affine gravity is the transition from nonminimal to minimal couplings which is realized by only field redefinition of the scalar fields. Consequently, the inflationary models gain a unique description in this context where the observed parameters, like the scalar tilt and the tensor-to-scalar ratio, are invariant under field reparametrization. Overall, gravity in its affine approach is expected to reveal interesting and rich phenomenology in cosmology and astroparticle physics. 
}

\begin{document}
\maketitle
\flushbottom

\section{Introductory remarks and motivation}
\label{sec:introduction}

Cosmological inflation manifests itself as the most appealing scenario for solving the problems of the big bang initial conditions, namely, the flatness and horizon problems \cite{guth, linde1, albrecht, linde2}. Furthermore, this early phase of rapid expansion is found to serve an excellent explanation to the observed cosmic microwave background anisotropies \cite{planck}. In the standard view, the inflaton, a hypothetical scalar field, drives cosmic inflation by slowly rolling down the potential energy which dominates the energy density of the universe at the early stage. This is the standard slow roll inflation where the so called the rate of inflation roll is small. However, in different models of inflation, this rate may remain constant leading to \textit{constant-roll} inflationary scenarios \cite{const-roll1,const-roll2}.

\begin{figure}[tbp]
 \centering
  \includegraphics[width=0.8\textwidth]{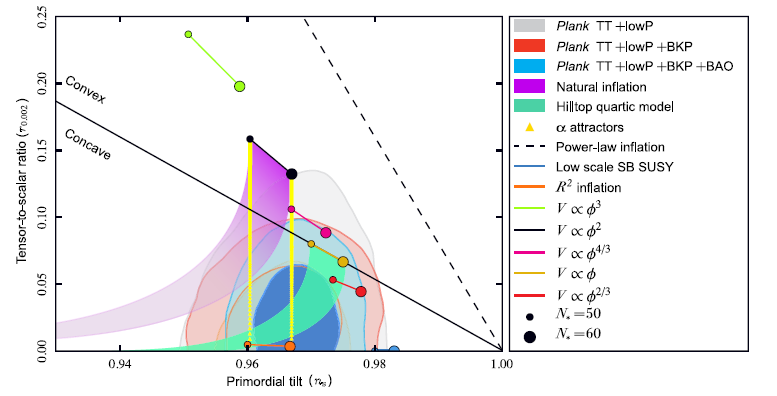}
  \caption{($n_{s},r$) bounds from Planck \cite{planck}. The results favor only models with small tensor-to-scalar ratio, this is the case of Starobinski model, natural inflation, $\alpha$-attractors and some other models. }
  \label{fig:planck results}
\end{figure}

Crucial prediction of inflation is the nearly scale-invariant cosmological scalar perturbations translated by the smallness (nearly unity) of the scalar tilt $n_{s}$. Additionally, generation of tensor perturbations is also possible in most of the inflationary models, where its smallness is provided by the so called tensor-to-scalar ratio $r$. Last few years accurate cosmological data has offered a powerful discrimination between different theories, and helped in supporting or ruling out various inflationary models (see Figure \ref{fig:planck results}.)

In the simplest inflationary models, the inflaton field is coupled minimally to Einstein's gravity and the slow roll conditions applied to the field impose some conditions on the potentials. Thus a good theory of inflation requires potentials that satisfy these constraints. Potentials are generally dependent on physical parameters like the field mass and its self coupling parameter, and in order to produce an amplitude of density perturbations required by the observational data, one has to severely fine-tune some of these physical parameters. Since this is unacceptable, it becomes difficult to support minimally coupled fields. The simplest way out to this problem is to keep the same field (inflaton) and go beyond minimal couplings to gravity. In the recent  years, nonminimal coupling to gravity becomes a subject of interest which is applied to both particle physics and cosmology. The key point is to add an interaction of the scalar field with the spacetime curvature which leads to a modified theory of gravity, this is usually called \enquote{scalar-curvature theory}\cite{steinhardt extended inf, barrow1, kalara, amendola, salopek, kolb, makino1, fakir1, fakir2, kaiser, komatsu,futamase, nozari}. If the nonminimal coupling parameter, noted $\xi$, is taken large enough, one may easily get a small enough density perturbations without adjusting any physical parameter. This is exactly what motivates the standard model (SM) Higgs boson as a possible candidate to drive inflation \cite{higgs inflation}.

Having a reasonable amplitude of density perturbation is not the only reason that motivates nonminimal couplings. In fact, it has been known for a long time that these sorts of couplings are no longer avoidable in quantum field theory in curved spacetime. It turned out that these terms are generated through the quantum corrections to the law energy effective actions by integrating out high energy degrees of freedom, thus, a nonzero parameter $\xi$ appears automatically at some energy scale even if it is zero at the tree level \cite{qft}. In this case, higher order curvature terms are not avoidable in curved spacetime for renormalization. For more details on how these terms appear through the effective action in curved spacetime, we refer the reader to the interesting review by Buchbinder et al \cite{effective action in qg}. Although the nonminimal coupling parameter $\xi$ is fixed depending on the cosmological models at hand, however, as it has been shown in \cite{bounds on xi} some bounds of this parameter arise precisely from quantum field theory in curved spacetime.      

The original action, written in \textit{Jordan frame}, where the nonminimal coupling terms appear explicitly can be transformed to \textit{Einstein frame} via \textit{conformal transformation}, where the nonminimal terms disappear. It is important to mention here that this transformation does not arise only in the nonminimal coupling to gravity but it also holds in F(R) gravity where the latter is mapped to Einstein's gravity in the presence of a scalar field \cite{f(R)}. 

In Einstein frame, the theory is brought to a new (redefined) scalar field  minimally coupled to gravity through the new metric\footnote{The new frame is named Einstein frame since the theory is brought to Einstein's General Relativity (GR) plus matter. This will be clear in the next section.}. Since one frame can be recast to another through this transformation, the frames must be equivalent. Inequalities of these frames, however, arise in different cosmological and particle physics applications \cite{magnano, faraoni, damour1, maeda conformal transformation, gottlober, barrow2, damour2, cotsakis, bekenstein}. The question then arises whether \textit{Jordan} or \textit{Einstein} frame represents the \textit{physical frame}.

In what follows, we enumerate some of the important physical cases where conformal frames possibly lead to ambiguities:

\begin{enumerate}
\item \textbf{Classical gravity and cosmology}

It was argued that even at the classical level, the Jordan frame is not physical and that the theory formulated in this frame makes sense only if one can define a transformation that recast the original theory to general relativity (Einstein frame) \cite{magnano}. The physical reason that led the authors to this conclusion is the indefinite sign of the energy which unlike in Einstein frame, it may possibly lead to unacceptable negative energy density in Jordan frame. In contrast to this, some believe that classically, the two frames are physically equivalent once the units of the observed quantities and the fundamental parameters, like mass, length and time are scaled with the conformal factor in Einstein frame \cite{equivalence of the frames}.

Another issue is the violation of Einstein's equivalence principle, this arises when massive fermions and gauge bosons are present in the theory in addition to the scalar field which is nonminimally coupled to gravity in Jordan frame. In Einstein frame, massive fields which are not invariant under conformal transformation will have a natural interaction with the scalar field, this means that these fields will not follow the geodesics of the present frame due to the appearance of an additional force as a result of the conformal transformation. However, massless particles keep following null geodesics in Einstein frame.

When it comes to cosmology, it has been shown that if an accelerated phase is imposed in one of the conformal frames, there is no guarantee to have an accelerating phase in the other frame when the conformal transformation is applied, the same ambiguity holds when treating singularities \cite{f(R)}.

In inflationary cosmology, the calculation of the spectral index in various frames shows significant differences. The ambiguity here may be resulted from different parameters written in various frames, such as the slow-roll parameters and the number of $e$-folds which are generally not invariant under conformal transformation. But in general, the problem is traced back to the calculation of the power spectrum of the cosmological perturbation where the procedure is not purely classical (semi-classical calculation is relevant.) Here we have to mention that frame ambiguities where different results are seen in different frames arise also in $\alpha$-attractor inflationary models \cite{frame-alpha-attractor}.

\item \textbf{Quantum fluctuations and corrections to the physical parameters}

Even if it happens that the frames could be physically equivalent classically, the equivalence certainly breaks down at the quantum level. This is the case when we quantize the scalar fields in curved background where corrections to all physical parameters (including the parameter $\xi$) arise naturally through the effective potential \cite{effective potential}. 

Power spectrum of the density perturbations is generally derived after computing the two point correlation function. This latter is related to the quantum perturbation of the inflaton field and the vacuum state. In the two conformal frames, the quantum fluctuations of the inflaton (the original field and the rescaled one) are not equivalent, furthermore, the vacuum state chosen in one frame is no longer the same state when it is mapped into the other frame. These ambiguities, generally, lead to two different values of the spectral index $n_{s}$ which is the important observable quantity for any model of inflation. However, it has been shown that the calculations in the two frames may lead to an equivalent results for some special cases, like chaotic inflation, where the expansion rate of the scale factor is quasi-de Sitter \cite{sasaki,kaiser2}.

SM Higgs boson provides an excellent classical background for inflation where the predicted spectrum of scalar and tensor perturbations are in accordance with observation. However, it has been shown that for nonminimally coupled Higgs boson, the quantum effects become essential. In fact, the scale of inflation is some orders of magnitude higher than the electroweak scale, and then, the running of the coupling constants from the electroweak scale (where the SM couplings are measured) to the inflationary scale becomes significant. It turned out that the quantum corrections maintained in Einstein and Jordan frames are inequivalent \cite{steinwachs}.

At the quantum level, the debate may not settle down without a complete theory of quantum gravity.
\end{enumerate}

Conformal frames are generally described by both, \textit{metric} and \textit{field} reparametrizations. The first is introduced to recast the gravitational part of the action to Einstein-Hilbert action, whereas the second transformation brings a canonical kinetic term of matter. If the theory describes a single scalar field, the canonical kinetic term is easily obtained after the transformations, however, if multifields are nonminimally coupled to gravity, the transition to Einstein frame with canonical kinetic terms of all the fields is not trivial, and generally is impossible. The reason of this difficulty is the derivative (of the fields) terms that appear in the new transformed curvature and contribute to the kinetic parts of the fields.

The ambiguities then may emerge from the redefinition of the \enquote{metric} tensor and not matter. The latter may be redefined and enter the calculations as a new variable without altering the physics, however, the new (conformal) metric tensor describes a \enquote{new} gravitational field. In fact, the metric transformation is performed in the same coordinate system of the spacetime manifold and it is not associated with a diffeomorphism of the manifold where the \enquote{transformed metrics} represent the same gravitational field.

The question that arises now is whether instead of using the conformal transformation, the transition between the couplings may be performed only through some \enquote{special} field redefinitions without altering the gravitational sector. This is indeed impossible if the gravitational theory is \textit{purely metric} like GR, however, a possible non-metric theory of gravity where the metric tensor is not postulated \textit{a priori} may serve a way out to avoid the use of conformal frames, this is the case of purely Affine Gravity (AG). In AG, matter (scalar field) must interact with gravity through affine connection and curvature rather than metric, the spacetime background then does not recognize any metric structure and it is only through dynamical equations that the metric tensor appears. Generating the metric tensor in this way provides an origin to the frame itself which must be unique.

In this work, we review a general framework of scalar fields interacting with \enquote{affine} gravity and tackle the problem of conformal frames from this side. We argue that conformal frames based on conformal transformations are not present in the new setup. This can be understood from the fact that affine gravity provides an origin to the metric elasticity of space, which is encoded in the metric tensor, and this latter is unique. In metric gravity like GR, the metric tensor resides in the spacetime \textit{a priori} and then conformal transformations are performed even before deriving the equations of motion, however, the metric tensor of affine gravity is the result of the equation of motion and it describes the \enquote{physical} and unique gravitational field. The transformation needed in the action is performed only through field redefinition (metric is absent there.) A remarkable consequence of this feature is the unique description of the inflationary models in the context of affine gravity \cite{affine inflation,induced affine inflation}.  

We organize the present paper as follows: In section \ref{sec:conformal frames in metric gravity}, we discuss briefly the nonminimal coupling in metric gravity by providing both physical and mathematical constructions (this is discussed in more details by several authors in the literatures which we refer here.) We show how to recast the theory to Einstein frame using the conformal transformation and field redefinition in the case of a single scalar field, and we illustrate how this transformation fails to bring a canonical kinetic terms if multifields are present in the nonminimal coupling terms. In the same section, we apply the field equations to slow-roll inflation and study the transition between the conformal frames by paying attention to the frame dependent quantities that lead to the ambiguities. Among these are the number of $e$-foldings and especially the intrinsic curvature perturbation. Section \ref{sec: field re-parametrization} will be devoted to the general framework of pure affine gravity. We will review minimal and nonminimal couplings to affine gravity and show how the transition between the two is made via the scalar field redefinition, even for the case of multifield dynamics. An important consequence of this property is the invariance of the intrinsic curvature perturbation. 

In section \ref{sec: affine gravity induces metric gravity}, we present the affine approach to induced gravity. This important part shows how pure affine gravity is able to induce both, the gravitational scale and metric elasticity of space from the very existence of spacetime, which is endowed with an affine connection and a heavy scalar field with nonzero vacuum energy. While vacuum expectation value of the scalar field leads to the Planck mass, the nonzero vacuum energy plays an important role in inducing the metric tensor. 
 
In section \ref{sec:inflationary models} we discuss how to realize inflationary models in the context of affine gravity by studying some of the known models. In section \ref{sec:conclusion} we summarize and finally we provide the reader with two appendices that include the conformal transformation and especially the affine dynamics which we suppose that the reader is not familiar with.

\section{Conformal frames in metric gravities}
\label{sec:conformal frames in metric gravity}
\subsection{Single scalar field}

In the context of metric gravity, the gravitational field is described by the metric tensor $g_{\mu\nu}$, this latter is essential in the theory and it generalizes the Minkowski metric in flat spacetime. In this case, we say that the spacetime manifold is endowed with a Lorentzian metric tensor. Matter then can be coupled to gravity directly via this field (metric) and then we say that the coupling is minimal. This coupling has its origin from the equivalence principle where gravity is incorporated in our actions by transforming the Minkowski metric to a general (curved) metric. However, it turns out that for different reasons (at least at the quantum level) matter may also be coupled to gravity through the curvature of spacetime. The obtained interaction is called nonminimal coupling.

The simplest form of the interaction of a single scalar field $\phi$ with both metric and curvature is given by the following action
\begin{eqnarray}
\label{non-minimal coupling metric action}
S\left[g, \phi \right]
=\int d^{4}x \sqrt{||g||} \left[\frac{M_{Pl}^{2}}{2}R\left(g\right) -\frac{1}{2}g^{\mu\nu}\nabla_{\mu}\phi \nabla_{\nu}\phi -V\left(\phi\right) +\frac{\xi\phi^{2}}{2}R\left(g\right)  \right],
\end{eqnarray}
where the sign ||.|| refers to the absolute value of the determinant, and $\xi$ is a dimensionless constant.

The theory described by the last action is said to be formulated in Jordan frame\footnote{Here, the kinetic term of the filed $\phi$ is written in a canonical form, however, this can be generalized to non canonical terms such as
$\psi(\phi)g^{\mu\nu}\nabla_{\mu}\phi \nabla_{\nu}\phi$. These models of scalar-tensor gravity are called Brans-Dicke theories \cite{jordan,brans}. }.

Action (\ref{non-minimal coupling metric action}) is varied with respect to the metric tensor and the scalar field leading to the coupled field equations
\begin{eqnarray}
\label{non-minimal coupling metric gravitational equations}
G_{\mu\nu}\left(g\right)=
\frac{1}{M_{Pl}^{2}+\xi \phi^{2}}\left[ \nabla_{\mu}\phi \nabla_{\nu}\phi
-\frac{1}{2}g_{\mu\nu}(\nabla \phi)^{2}-g_{\mu\nu}V\left(\phi\right)+\xi \nabla_{\mu}\nabla_{\nu}\phi^{2}-\xi \Box \phi^{2}g_{\mu\nu} \right]
\end{eqnarray}
and
\begin{eqnarray}
\label{non-minimal coupling metric phi equation}
\Box \phi -V^{\prime}\left(\phi \right)+\xi\phi R\left(g\right)=0,
\end{eqnarray}
where $G_{\mu\nu}\left(g\right)=R_{\mu\nu}-\frac{1}{2} g_{\mu\nu}R$ is the Einstein tensor.

It is clear that the theory coincides with Einstein gravity (Einstein-Hilbert action plus matter) for a zero nonminimal coupling parameter $\xi=0$. In the general case, where $\xi \neq 0$, it is always possible to write the previous theory (the action) in a form that looks equivalent to Einstein gravity in which the coupling matter-gravity is minimal. This transition requires the so called \enquote{conformal transformation}.

A conformal transformation allows the passage from one metric tensor $g_{\mu\nu}$ to another $\tilde{g}_{\mu\nu}$ in the same spacetime coordinates by
\begin{eqnarray}
\label{conformal transformation}
\tilde{g}_{\mu\nu}=\mathcal{F}\left(\phi\right)g_{\mu\nu},
\end{eqnarray}
where $\mathcal{F}\left(\phi\right)$ is a general smooth function which in our case, takes the form
\begin{eqnarray}
\mathcal{F}\left(\phi\right)=1+\frac{\xi \phi^{2}}{M^{2}_{Pl}}.
\end{eqnarray}
This function may take a general form if the coupling term at the end of the action (\ref{non-minimal coupling metric action}) is an arbitrary function of $\phi$, and then, all the properties that we will study here are also applied to the general case.

Although it brings the gravitational sector to its Einstein-Hilbert form, the above transformation generates additional terms which break the canonical form of the kinetic terms of the matter field (see appendix \ref{appendix1}). This obliges us to redefine the field $\phi$ and its potential as follows
\begin{eqnarray}
\label{field redefinition}
d\tilde{\phi} = \sqrt{\frac{1}{\mathcal{F}\left(\phi\right)}+\frac{3\mathcal{F}^{\prime 2}\left(\phi\right)}{2 M^{2}_{Pl}\mathcal{F}^{2}\left(\phi\right) } }
\,\, d\phi,
\end{eqnarray}
with a potential
\begin{eqnarray}
\label{potential redefinition}
\tilde{V}(\tilde{\phi})=\frac{V\left(\phi\right)}{\mathcal{F}^{2}\left(\phi\right)}.
\end{eqnarray}
Finally, with these transformations, the action (\ref{non-minimal coupling metric action}) is simply brought to the following Einstein-Hilbert action with a scalar field $\tilde{\phi}$ minimally coupled to the new metric $\tilde{g}_{\mu\nu}$
\begin{eqnarray}
\label{minimal coupling metric action}
S\left[g, \phi \right] \rightarrow
\tilde{S}[\tilde{g}, \tilde{\phi} ]
=\int d^{4}x \sqrt{-\tilde{g}} \left[\frac{M_{Pl}^{2}}{2}\tilde{R}\left(\tilde{g}\right) -\frac{1}{2}\tilde{g}^{\mu\nu} \tilde{\nabla}_{\mu}\tilde{\phi}
\tilde{\nabla}_{\nu}\tilde{\phi} -\tilde{V}(\tilde{\phi}) \right],
\end{eqnarray}
where the tilde refers to the transformed quantities.
 
It is for this reason we say that the theory now, is formulated in Einstein frame. Clearly, the Jordan frame $\left(\phi,g\right)$ is conformally transformed to Einstein frame $(\tilde{\phi},\tilde{g})$ and vice versa.

It is worth noting that the conformal transformation (\ref{conformal transformation}) is not a diffeomerphism of the spacetime manifold, and then, the new metric $\tilde{g}_{\mu\nu}$ is not the original metric $g_{\mu\nu}$ seen by a different observer. Therein lies the problem of the conformal frames, in fact, the two metrics describe two different gravitational fields and the two frames then may describe different physics. The question that arises now is which one of these frames describes the reality ? After all, and if the two frames predict different results, it is only one of them that might be confronted with observations.

The problems with these frames have been discussed in much details by several authors from the classical view, like the violation of the weak energy conditions in Jordan frame, and the violation of the equivalence principle in Einstein frame where the scalar field becomes part of the metric tensor (the gravitational field)\cite{magnano, faraoni}. Serious ambiguities are discussed from quantum mechanical view where quantum corrections to both matter and gravity are no longer compatible in the two frames \cite{dilatonic,steinwachs}. Here, the quantum fluctuations in the two frames refer to different metric tensors. Since the new metric tensor (Einstein-frame) includes the scalar degree of freedom, the latter is automatically mixed with tensor modes. This fact would lead to difficulties of obtaining the same results, when these latter are transformed back to the original frame.

\subsection{Multifield case}
The nonminimal coupling to gravity holds even for mutifield models. In this case the function $\mathcal{F}$ becomes a general function of the fields $\phi^{A}=\phi^{1}, \dots,\phi^{N}$, and the invariant action takes the following form
\begin{eqnarray}
\label{metric mutifield action}
S[g, \phi^{A} ]
=\int d^{4}x \sqrt{||g||} \left[\mathcal{F}(\phi^{1},\dots,\phi^{N}) R\left(g\right) -\frac{1}{2}
\delta_{AB} g^{\mu\nu}\nabla_{\mu}\phi^{A} \nabla_{\nu}\phi^{B} -V(\phi^{1},\dots,\phi^{N})  \right] \nonumber \\
\end{eqnarray}
where $N$ is the dimension of the field space.

This action is written in Jordan frame where the coupling matter-curvature appears explicitly. Here, the internal indices $A,B$ are raised and lowered by the flat (Euclidian) metric $\delta_{AB}$ of the field space.

To recast this action to Einstein frame, the metric tensor must be conformally transformed to a new metric $\tilde{g}_{\mu\nu}$ as
\begin{eqnarray}
\label{multifield conformal transformation}
g_{\mu\nu} \rightarrow \frac{2}{M_{Pl}^{2}}
\mathcal{F}(\phi^{1},\dots,\phi^{N})
\tilde{g}_{\mu\nu}.
\end{eqnarray}
Now, the action (\ref{metric mutifield action}) takes the form
\begin{eqnarray}
\label{einstein frame mutifield}
S\left[g, \phi^{A} \right] \rightarrow
\int d^{4}x \sqrt{||\tilde{g}||}\left[\frac{M^{2}_{Pl}}{2} \tilde{R}(\tilde{g})
-\frac{1}{2}M_{AB}\tilde{g}^{\mu\nu}\tilde{\nabla}_{\mu}\phi^{A} \tilde{\nabla}_{\nu}\phi^{B} -\tilde{V}(\phi^{1},\dots,\phi^{N}) \right],
\end{eqnarray}
where the potential energy is written in terms of the Jordan frame potential as
\begin{eqnarray}
\tilde{V}(\phi^{1},\dots,\phi^{N})=
\frac{M^{4}_{Pl}}{4 \mathcal{F}^{2}(\phi^{1},\dots,\phi^{N})}
V(\phi^{1},\dots,\phi^{N}).
\end{eqnarray}
The operator $M_{AB}$ has the following form
\begin{eqnarray}
\label{matrix m-metric gravity}
M_{AB}=\frac{M^{2}_{Pl}}{2\mathcal{F}}
\left[\delta_{AB}+\frac{3}{\mathcal{F}}\frac{\partial \mathcal{F}}{\partial \phi^{A}}
\frac{\partial \mathcal{F}}{\partial \phi^{B}}  \right].
\end{eqnarray}
This quantity defines a metric tensor in field space which is impossible to be reduced to the flat metric, i.e, $\delta_{AB}$, for general field space dimension $N$. The reason for this is the second term in (\ref{matrix m-metric gravity}). A flat metric is obtained only if all the components of the Riemann curvature tensor constructed from the metric $M_{AB}$ vanish, this condition is not valid for dimensions $N>2$. This shows that the matter part in action (\ref{einstein frame mutifield}) can not be brought into its canonical form through any rescaling. Then, even in Einstein frame, the gravitational and matter sectors are not written together in their canonical forms. It has been noted that there are law-energy regime where the transformed action relaxes towards canonical form up to corrections that scale as $\xi_{A}^{2}(\phi^{A})^{2}/M^{2}_{Pl}$, where $\xi_{A}$ is the nonminimal coupling parameter that corresponds to the field $\phi^{A}$ \cite{kaiser conformal transformation}.

The difficulty of getting a canonical term stems not from field redefinition, but it is due to the metric conformal mapping itself. It is the transformation of the Ricci scalar (under this mapping) that brings the additional kinetic terms. This shows a first possible difficulty of using conformal transformation. In the following subsection, we will study the slow roll parameters required for inflation, in both, Jordan and Einstein frames, and then see how they transform under conformal transformation, leading to some ambiguities.

\subsection{Slow-roll inflation in different frames}

We are now in a position to see how the ambiguity of conformal frames occurs in inflation. In most cases, inflationary dynamics is studied using the standard slow-roll conditions applied to the inflaton, this latter rolls down the flat potential energy which dominates the total energy density of the universe. Let us now see how the slow-roll conditions are written and applied in Einstein and Jordan frames, and then make the transition between the two frames.
\newpage
\begin{enumerate}
\item \textit{Jordan frame}:

We choose for simplicity a single field (inflaton) $\phi$, and then the coupled field equations are given by equations (\ref{non-minimal coupling metric gravitational equations}) and (\ref{non-minimal coupling metric phi equation}). The flat Robertson-Walker line element is written in this frame in terms of the scale factor $a(t)$ as

\begin{eqnarray}
ds^{2}=-dt^{2}+ a^{2}(t)\delta_{ij}dx^{i}dx^{j}.
\end{eqnarray}
In this frame, the slow-roll conditions are written in their standard form as follows
\begin{eqnarray}
\label{slow-roll conditions in jordan frame}
\left|\frac{\ddot{\phi}}{\dot{\phi}} \right| \ll H \, ,
\quad \quad
\left|\frac{\dot{\phi}}{\dot{\phi}} \right| \ll H \, ,
\quad \quad
\dot{\phi}^{2} \ll V(\phi) \, ,
\quad \quad
\left|\dot{H} \right| \ll H^{2},
\end{eqnarray}
where $H$ is the Hubble parameter.

This simplifies the equations of motion of the background field $\phi(t)$ as
\begin{eqnarray}
H^{2}\simeq 
\frac{1}{3M^{2}_{Pl}\mathcal{F}(\phi)}
\left\lbrace
V(\phi)-\frac{M^{2}_{Pl}\mathcal{F}^{\prime}(\phi)}{1+(\mathcal{F}(\phi)-1)(1+6\xi)}
\left[2\mathcal{F}^{\prime}(\phi)V(\phi)-\mathcal{F}(\phi)V^{\prime}(\phi)
\right]
\right\rbrace \nonumber \\
\,
\end{eqnarray}
and
\begin{eqnarray}
3 H \dot{\phi} \simeq 
\frac{1}{1+(\mathcal{F}(\phi)-1)(1+6\xi)}
\left\lbrace
2\mathcal{F}^{\prime}(\phi)V(\phi)-\mathcal{F}(\phi)V^{\prime}(\phi)
\right\rbrace,
\end{eqnarray}
where 
\begin{eqnarray}
\mathcal{F}(\phi)=1+\frac{\xi \phi^{2}}{M^{2}_{Pl}}.
\end{eqnarray}
The slow-roll conditions are satisfied until the end of the inflationary phase, and they can be expressed using the slow-roll parameters defined by
\begin{eqnarray}
\epsilon \equiv -\frac{\dot{H}}{H^{2}} \,
\quad \quad
\eta \equiv -\frac{\ddot{H}}{H\dot{H}},
\end{eqnarray}
which remain less than unity during the inflationary regime.

Inflation then ends at the field $\phi_{end}$ which corresponds to $\epsilon=1$. Finding $\phi_{end}$ may not be quite difficult for general potentials, however, solving for the value of the field $\phi_{start}$, that corresponds to the time of horizon crossing, is not trivial. Nevertheless, the difficulty may be surmounted by using the number of $e$-folds N, where the scales of interest crossed outside of the horizon almost $N_{\star}=62$ $e$-folds before inflation ends. The number of $e$-foldings is generally defined as
\begin{eqnarray}
 N && \equiv \int_{t_{start}}^{t_{end}} H dt 
= \int_{\phi_{start}}^{\phi_{end}}
\frac{H}{\dot{\phi}} d \phi.
\end{eqnarray}
In practice, the integrand of this expression will be expressed in terms of the field $\phi$ only, via the potential and its first derivative. At first order, the spectral index $n_{s}$ is obtained then by evaluating the slow-roll parameters at the field $\phi_{start}$ leaving only a dependence on $N_{\star}$. This direct procedure of calculating the scalar tilt, shows that this latter may remain invariant if the field $\phi$ is naively redefined. However, as we will see later, this is no longer the case if the metric is also redefined.

\item \textit{Einstein frame}:

This frame serves a simple area for studying the inflationary dynamics. The gravitational field equations are nothing but Einstein equations plus scalar field. The equations of motion are simply derived from the transformed action (\ref{minimal coupling metric action}) by varying with respect to the metric $\tilde{g}^{\mu\nu}$ and the field $\tilde{\phi}$.

Writing the flat Robertson-Walker metric in this frame as
\begin{eqnarray}
d\tilde{s}^{2}=
-d\tilde{t}^{2}+ \tilde{a}^{2}(\tilde{t})\delta_{ij}dx^{i}dx^{j},
\end{eqnarray}
and then the cosmological dynamics is governed by the coupled equations
\begin{eqnarray}
\label{cosmological dynamics in einstein frame1}
\tilde{H}^{2}=\frac{1}{3 M^{2}_{Pl}}\left(\frac{\dot{\tilde{\phi}}^{2}}{2}+\tilde{V}(\tilde{\phi}) \right),
\end{eqnarray}
\begin{eqnarray}
\label{cosmological dynamics in einstein frame2}
\ddot{\tilde{\phi}}+ 3\tilde{H}\dot{\tilde{\phi}}+\tilde{V}^{\prime}(\tilde{\phi})=0.
\end{eqnarray}
Here the Hubble parameter is written as $\tilde{H}=\dot{\tilde{a}}/\tilde{a}$, where the dot refers to the derivative with respect to $\tilde{t}$, and the prime denotes the derivative with respect to $\tilde{\phi}$. The spacial coordinates $x^{i}$ are not subjected to conformal transformation.

In Einstein frame, the slow-roll conditions are given by
\begin{eqnarray}
\label{slow-roll conditions in einstein frame}
\dot{\tilde{\phi}}^{2} \ll \tilde{V}^{\prime} \, ,
\quad \quad \left|\frac{\ddot{\tilde{\phi}}}{\dot{\tilde{\phi}}} \right| \ll \left| \tilde{H}
\right|,
\end{eqnarray}
which simplify the equations of motion (\ref{cosmological dynamics in einstein frame1}) and (\ref{cosmological dynamics in einstein frame2}).

The slow-roll parameters are written here in terms of the potential and its derivative
\begin{eqnarray}
\tilde{\epsilon}=\frac{M^{2}_{Pl}}{2}\left(\frac{\tilde{V}^{\prime}}{\tilde{V}} \right)^{2} \,
,\quad \quad \tilde{\eta}= M^{2}_{Pl}\frac{\tilde{V}^{\prime \prime}}{\tilde{V}} \,
,\quad \quad
\tilde{\zeta}^{2}=M^{4}_{Pl}\frac{\tilde{V}^{\prime}\tilde{V}^{\prime \prime \prime}}{\tilde{V}^{2}}.
\end{eqnarray}
The same procedure followed in Jordan frame is applied here to calculation of the values of the fields $\tilde{\phi}_{start}$ and $\tilde{\phi}_{end}$ that correspond to the horizon crossing and the end of inflation respectively, in this frame.

In this case, the number of $e$-foldings $\tilde{N}$ takes the form
\begin{eqnarray}
\tilde{N}\equiv  \int_{\tilde{t}_{start}}^{\tilde{t}_{end}} \tilde{H}
d \tilde{t}
= -\frac{1}{M^{2}_{Pl}}
\int_{\tilde{\phi}_{start}}^{\tilde{\phi}_{end}}
\frac{\tilde{V}}{\tilde{V}^{\prime}} d \tilde{\phi}.
\end{eqnarray}
Again, at first order, the scalar tilt in this frame, $\tilde{n}_{s}=1-6\tilde{\epsilon}+2\tilde{\eta}$ can be obtained in terms of $\tilde{N}_{\star}=62$.

Einstein frame is usually the preferred frame to study inflation. In fact, it is in this frame that firstly we write the intrinsic curvature perturbation, this is essential in calculating the spectrum of the density perturbation. During inflation, the intrinsic curvature perturbation is given by \cite{liddle,stewart}
\begin{eqnarray}
\label{metric intrinsic curvature perturbation}
\tilde{\mathcal{R}}= \frac{\tilde{H}}{\dot{\tilde{\phi}}} \delta \tilde{\phi},
\end{eqnarray}
where $\delta \tilde{\phi}$ denotes the quantum fluctuation of the inflaton field $\tilde{\phi}$.

The slow-roll formalism is based on the calculation of the spectrum of the density perturbation
\begin{eqnarray}
\tilde{\mathcal{P}}_{\tilde{\mathcal{R}}}^{1/2}=
\frac{\tilde{H}}{\dot{\tilde{\phi}}} \sqrt{\mid \Delta \tilde{\phi} \mid^{2}},
\end{eqnarray}
where $\mid \Delta \tilde{\phi} \mid^{2}$ is the two point correlation function for $\delta \tilde{\phi}$.
The scalar spectral index is defined through this basic quantity by
\begin{eqnarray}
\tilde{n}_{s}-1 \equiv \frac{d \ln \tilde{\mathcal{P}}_{\tilde{\mathcal{R}}} }
{d \ln k},
\end{eqnarray}
with $k$ being the momentum which appears in Fourier transformation of the field $\delta\tilde{\phi}$.

As we will see below, the intrinsic curvature perturbation in its standard form (\ref{metric intrinsic curvature perturbation}) is not invariant under conformal transformation, which then makes an ambiguity when performing the calculation in Jordan frame.

\item \textit{Transition between the frames and ambiguities}:

The cosmological parameters written in the two frames separately may now be mapped from one frame to another via the conformal transformation (\ref{conformal transformation}), and field redefinition (\ref{field redefinition}). To that end, the flat Robertson-Walker line element in Einstein frame becomes
\begin{eqnarray}
d\tilde{s}^{2}=
-d\tilde{t}^{2}+ \tilde{a}^{2}(\tilde{t})\delta_{ij}dx^{i}dx^{j} 
= \mathcal{F}\left(\phi \right)\left[-dt^{2}+ a^{2}(t)\delta_{ij}dx^{i}dx^{j} \right].
\end{eqnarray}
This leads to the relation between the cosmic times and the scalar factors in the two frames
\begin{eqnarray}
\label{cosmic time relation}
d \tilde{t}=\sqrt{\mathcal{F}} \,\, dt, \quad \quad \tilde{a} \left(\tilde{t}\right)
=\sqrt{\mathcal{F}}\,\, a\left(t\right).
\end{eqnarray}
Now, the Hubble parameter in Einstein frame becomes
\begin{eqnarray}
\label{hubble relation}
\tilde{H}=\frac{1}{\tilde{a}}\frac{d \tilde{a}}{d \tilde{t}}
=\frac{1}{\sqrt{\mathcal{F}}}\left(H+ \frac{\dot{\mathcal{F}}}{2 \mathcal{F}} \right).
\end{eqnarray}
This means that the \enquote{expansion} rate is not invariant and then the space patches expand differently in different frames. A first and direct consequence of this property is that the number of $e$-folds is frame dependent, in fact, using relations (\ref{cosmic time relation}) and (\ref{hubble relation}) we easily get
\begin{eqnarray}
\tilde{N}=  \int_{\tilde{t}_{start}}^{\tilde{t}_{end}} \tilde{H}
d \tilde{t} 
= N+\frac{1}{2} \ln \left(\frac{\mathcal{F}_{end}}{\mathcal{F}_{start}} \right).
\end{eqnarray}
The last term does not vanish in general, leaving a trivial \enquote{unwanted} contribution.
The problem that arises due to the frame dependence of the number of the $e$-foldings
is the fact that, if the scalar spectral indices were to be calculated by evaluating the slow-roll parameters in terms of $N$, then there would be no guarantee for getting the same value in both frames. This is what we face in practice in effect, it has been shown that for chaotic inflation, the scalar tilts calculated in Einstein and Jordan frames differ at second order \cite{kaiser,nozari}. The differences become significant when the expansion rate follows a power law, like the case of induced gravity inflation (see section \ref{sec:inflationary models} and Ref.\cite{kaiser}.)

As we have mentioned earlier, the origin of this slow-roll formalism is traced back to the form of the intrinsic curvature perturbation (\ref{metric intrinsic curvature perturbation}). The mean problem is that although this quantity is \enquote{gauge} invariant, it is not invariant under conformal transformation.

This is easily seen at first order of $\delta \phi$ as follows
\begin{eqnarray}
\label{transformed metric intrinsic curvature perturbation}
\tilde{\mathcal{R}}\equiv  \frac{\tilde{H}}{\dot{\tilde{\phi}}} \delta \tilde{\phi}
= \frac{1}{\sqrt{\mathcal{F}}}\left(H+ \frac{\dot{\mathcal{F}}}{2 \mathcal{F}} \right)\frac{\delta \phi}{\dot{\phi}} 
\neq  \mathcal{R},
\end{eqnarray}
where we have used the field redefinition (\ref{field redefinition}) and the transformations (\ref{cosmic time relation}) and (\ref{hubble relation}).
\end{enumerate}
The conclusion of all this is that although the conformal frames are \enquote{mathematically} equivalent, they lead to ambiguities in practice. It is worth noting however that in some models of inflation, the predictions in the two frames may be equivalent \cite{sasaki,kaiser2}.

In the next sections, we will explore the pure affine approach to gravity, and then discuss the transitions from nonminimal to minimal couplings in the affine context where the previous discussed parameters remain invariant. 

\section{Affine gravity: conformal frames or field redefinition?}
\label{sec: field re-parametrization}

\subsection{Single scalar field in affine space}
\label{subsec:single field models}
So far in this article, we have raised the question of whether the transition from nonminimal to minimal couplings to gravity may be performed by only field redefinition without altering the geometric part of the action. If the theory of gravity at hand is metric, this becomes difficult to achieve.

Metric theories of gravity, like GR, are based on the concept of metric. This concept is additional and it is not required in general curved spaces. The metrical structure is postulated in the spacetime since it provides us with the measurements of distances and angles which are encoded in the metric tensor. Although this latter is not avoidable at large scales, it maybe possible that this structure has been arisen and emerged \textit{a posteriori}, and that spacetime has started with a completely different and simpler structure.   

In the absence of the metric tensor, one may simply think about \textit{affine space}. This space is trivially endowed with an arbitrary \textit{affine connection} that provides the concept of parallel displacements and leads to a covariant comparison of tensors at different points in spacetime. Straight lines in this space are nothing but \textit{geodesics} of the geometry, not to extremize lengths, but to \textit{parallel transport} the tangent vectors. The curvature of spacetime in this case is measured through the geodesic deviations of test particles, and this leads to the concept of gravitational force. What is known as GR with its metrical structure can be simply generated from this simple affine structure.  

To formulate an affine theory of gravity, we need an affine connection and its associated curvature. This connection can be considered arbitrary, however, for simplicity it can be taken symmetric $\Gamma_{\nu\mu}^{\lambda}=\Gamma_{\mu\nu}^{\lambda}$. Then we proceed by defining the following quantities:
\begin{enumerate}
\item \textit{Invariant volume measure}:

This is important for getting a covariant equations of motion via the principle of least action, and it replaces the volume measure $\sqrt{||g||}$ of GR and other metric theories. A simple alternative is the square root of the determinant of a rank-two tensor. In affine space, this can be constructed from curvature, thus, the Ricci tensor $R_{\mu\nu}(\Gamma)$. If matter is plunged into the space as a simple scalar field $\phi$, then its kinetic structure $\nabla_{\mu}\phi\nabla_{\nu}\phi$ can play a good role in forming this invariant. Thus, the possible invariant volume measure will be considered as the square root of the determinant of the linear combination of both quantities; Ricci tensor and kinetic structure of the scalar field. For simplicity, we will be interested only in the symmetric part of the Ricci tensor, $R_{\mu\nu}=R_{(\mu\nu)}$.
\item \textit{Scalar integrand}:

The scalar field $\phi$ enters affine space through its kinetic structure, and it remains its potential energy $V(\phi)$. This is considered as any scalar function and it simply enters the action as a multiplicative term. However a special attention should be given to this part. As a multiplicative term, the case $V(\phi)=0$ would lead to zero or an infinite (singular) action. Both are unwanted and in order to avoid them, we must impose $V(\phi)\neq 0$ everywhere. This is a novel property which is restricted to affine gravity.      
\end{enumerate}  

Based on the properties stated above, we propose the following action
\begin{eqnarray}
\label{affine single0}
S\left[\Gamma,\phi \right] = \int d^{4}x \frac{\sqrt{ \left| \right| \left(M^2 + \xi \phi^2\right)R_{\mu\nu}\left(\Gamma\right) - \nabla_{\mu}\phi \nabla_{\nu}\phi \left| \right|}}{V(\phi)}, \nonumber  \\
\end{eqnarray}
where $M$ is an arbitrary constant of mass dimension.

As one may easily show, this action is invariant under general coordinate transformations. Additionally, the action may acquire other internal symmetries depending on the potential energy. For instance, the term inside the determinant has a $Z_{2}$ symmetry.

Now, since the fundamental field is the affine connection $\Gamma$, then the field equations must be resulted from variation of action (\ref{affine single0}) with respect to it. To that end, one gets the following dynamical equation (see appendix \ref{appendix2} for explicit derivation)
\begin{eqnarray}
\label{dynamical equation single field}
\nabla_{\alpha}\left\lbrace \left(M^{2}+\xi \phi^{2} \right) \frac{\sqrt{\left| \right|  K_{\mu\nu}\left(\phi\right)  \left| \right|}}{V\left(\phi\right)}\left(K^{-1}\right)^{\mu\nu}
\right\rbrace =0,
\end{eqnarray}
where $\nabla$ is the covariant derivative with respect to the affine connection, and the tensor $K_{\mu\nu}$ is given by
\begin{eqnarray}
\label{tensor k(Gamma,phi)}
K_{\mu\nu}\left(\phi\right)=
\left(M^2 + \xi \phi^2\right)R_{\mu\nu}\left(\Gamma\right) - \nabla_{\mu}\phi \nabla_{\nu}\phi.
\end{eqnarray}
The solution to this equation is provided by the existence of a rank-two symmetric tensor $g_{\mu\nu}$ which defines with its inverse $(g^{-1})^{\mu\nu}$, a constant scalar density satisfying 
\begin{eqnarray}
\label{solution of the dynamical equation}
\left(M^{2}+\xi \phi^{2} \right)\frac{\sqrt{\left| \right|  K_{\mu\nu}\left(\phi\right)  \left| \right|}}{V\left(\phi\right)}\left(K^{-1}\right)^{\mu\nu}
=\bar{M}^{2} \sqrt{\left| \right| g \left| \right|} \left( g^{-1}\right)^{\mu\nu},
\end{eqnarray}
where $\bar{M}$ now, is a constant of integration.

This implies that $\nabla_{\alpha}g_{\mu\nu}=0$, and then the affine connection is reduced to the Levi-Civita connection of the tensor $g_{\mu\nu}$
\begin{eqnarray}
\label{levi-civita connection}
\Gamma^{\lambda}_{\mu\nu} \rightarrow
\Gamma^{\lambda}_{\mu\nu}(g)=
\frac{1}{2}g^{\lambda\sigma}(\partial_{\mu}g_{\nu\sigma}
+\partial_{\nu}g_{\mu\sigma}-\partial_{\sigma}g_{\mu\nu}).
\end{eqnarray}
The new tensor $g_{\mu\nu}$ with its compatibility condition that leads to its associated connection (\ref{levi-civita connection}) plays then the role of a \textit{metric tensor}. This metric tensor is not postulated \textit{a priori} as in GR, but it arises dynamically from the affine structure. This approach provides a first argument towards the \enquote{emergence} of metrical elasticity of space which we will explore latter in this article.

Before proceeding to the scalar field dynamics, we should point out here an important point that concerns the Lorentzian signature of the generated metric. At first glance, one may notice that the metric tensor is given in terms of the affine connection and the scalar field as in (\ref{solution of the dynamical equation}). In imposing the physical signature, the solution to this dynamical equation must be taken such that the tensor $K_{\mu\nu}(\Gamma,\phi)$ defined by (\ref{tensor k(Gamma,phi)}), has one signature, say $(-,+,+,+)$ \cite{kijowski1}. 

Given the \textit{a posterior} metrical structure, the equations of motion now are nothing but the equality (\ref{solution of the dynamical equation}), which is written as
\begin{eqnarray}
\label{ricci equation of motion}
\left(M^{2}+\xi \phi^{2} \right)R_{\mu\nu}
- \nabla_{\mu}\phi \nabla_{\nu}\phi =
g_{\mu\nu}\left(\frac{\bar{M}^{2}}{M^{2}+\xi \phi^{2}} \right) V\left(\phi \right).
\end{eqnarray}
Contracting, raising and lowering the spacetime indices in the standard way can be performed using the metric tensor. Thus, the equation of motion (\ref{ricci equation of motion}) can be easily recast to a standard form as
\begin{eqnarray}
R_{\mu\nu}-\frac{1}{2}g_{\mu\nu}R=&&
\frac{1}{M^{2}+\xi \phi^{2}}\left[\nabla_{\mu}\phi \nabla_{\nu}\phi
-\frac{1}{2}g_{\mu\nu}(\nabla \phi)^{2}-g_{\mu\nu}V\left(\phi\right)  \right]\nonumber \\
&& +g_{\mu\nu} \frac{M^{2}-\bar{M}^{2}+\xi \phi^{2}}{\left(M^{2}+\xi \phi^{2} \right)^{2}} V\left(\phi\right)
\end{eqnarray}
For the case $\xi=0$, Einstein's field equations for minimal coupled scalar field implies that both constants $M$ and $\bar{M}$ must equal the Plank mass
\begin{eqnarray}
\bar{M}=M=M_{Pl}.
\end{eqnarray}
The last condition shows that a single scalar field $\phi$ is coupled to gravity through affine connection and its Ricci tensor via the following action \cite{affine inflation}
\begin{eqnarray}
\label{affine single}
S_{\text{AG}}\left[\Gamma,\phi\right] = \int d^{4}x \frac{\sqrt{ \left| \right| \left(M_{Pl}^2 + \xi \phi^2\right)R_{\mu\nu}\left(\Gamma\right) - \nabla_{\mu}\phi \nabla_{\nu}\phi \left| \right|}}{V(\phi)}.
\end{eqnarray}
Finally, the gravitational field equations derived from the action (\ref{affine single}) are written as
\begin{eqnarray}
\label{nonminimal-affine einstein equations}
R_{\mu\nu}-\frac{1}{2}g_{\mu\nu}R=&&
\frac{1}{M_{Pl}^{2}+\xi \phi^{2}}\left[\nabla_{\mu}\phi \nabla_{\nu}\phi
-\frac{1}{2}g_{\mu\nu}(\nabla \phi)^{2}-g_{\mu\nu}V\left(\phi\right)  \right]\nonumber \\
&& +g_{\mu\nu} \frac{\xi \phi^{2}}{\left(M_{Pl}^{2}+\xi \phi^{2} \right)^{2}} V\left(\phi\right)
\end{eqnarray}
Now variation of the action (\ref{affine single}) with respect to the scalar field $\phi$ leads to the following equation of motion
\begin{eqnarray}
\label{nonminimal-affine scalar field equation}
\Box \phi -V^{\prime}\left(\phi \right)+\xi \phi R\left(g\right)+\Psi\left(\phi\right)=0,
\end{eqnarray}
where the function $\Psi\left(\phi\right)$ is given by
\begin{eqnarray}
\label{Psi}
\Psi\left(\phi\right)=
\frac{\xi\phi^{2}}{M^{2}_{Pl}+\xi\phi^{2}}V^{\prime}\left(\phi\right) 
-\left(\frac{2\xi\phi}{M^{2}_{Pl}+\xi\phi^{2}}\right) g^{\mu\nu}\nabla_{\mu}\phi\nabla_{\nu}\phi.
\end{eqnarray}
In conclusion, we point out the following differences between Affine Gravity (AG) described by action (\ref{affine single}) and Metric Gravity (MG) based on action (\ref{non-minimal coupling metric action}):
\begin{enumerate}
\item The theories are conceptionally different since they are based on different fundamental fields. In MG, matter couples to the metric, whereas this latter is absent in AG, and matter then couples to affine connection.
\item Nevertheless, the theories provide equivalent equations of motion for the minimal coupling case.
\item The theories are inequivalent in the presence of nonminimal couplings.
\end{enumerate}

\subsection{Mapping to minimal coupling and invariant curvature perturbation }

The question now is how to recast the gravitational field equations (\ref{nonminimal-affine einstein equations}) to standard Einstein equations? What is the alternative to conformal transformation in this setup? The answer to this is that there is no need for conformal mapping to get the standard Einstein equation. In fact, one only needs to redefine the scalar field $\phi$ and its potential $V(\phi)$ as
\begin{eqnarray}
\label{affine field transformation}
d\tilde{\phi}=\frac{d\phi}{\sqrt{\mathcal{F}\left(\phi\right)}}, \quad \quad \text{and} \quad \tilde{V}[\tilde{\phi}(\phi)]=\frac{V(\phi)}{\mathcal{F}^{2}(\phi)}.
\end{eqnarray}
In terms of the new field $\tilde{\phi}$, one may easily show that the field equations (\ref{nonminimal-affine einstein equations}) and (\ref{nonminimal-affine scalar field equation}) are, respectively, written as
\begin{eqnarray}
\label{minimal-affine-grav-eq}
R_{\mu\nu}-\frac{1}{2}g_{\mu\nu}R=
M_{Pl}^{-2}\left[\nabla_{\mu}\tilde{\phi} \nabla_{\nu}\tilde{\phi}
-\frac{1}{2}g_{\mu\nu}(\nabla \tilde{\phi})^{2}-g_{\mu\nu}\tilde{V}(\tilde{\phi})  \right], 
\end{eqnarray}
\begin{eqnarray}
\Box \tilde{\phi}- \tilde{V}(\tilde{\phi})=0.
\label{minimal affine field equations}
\end{eqnarray}
These equations are familiar in general relativity, they describe the dynamics of a scalar field $\tilde{\phi}$ minimally coupled to gravity via the metric tensor $g_{\mu\nu}$. In other words, both fields are coupled (through equations of motion) to the same metric which is generated dynamically in our setup. This can be seen in a standard form from the transformation of the action (\ref{affine single}) under the field redefinition (\ref{affine field transformation})
\begin{eqnarray}
\label{transformed affine action}
S_{\text{AG}}\left[\Gamma,\phi\right] \rightarrow
\int d^{4}x \frac{\sqrt{ \left| \right| M_{Pl}^2 R_{\mu\nu}\left(\Gamma\right) - \nabla_{\mu}\tilde{\phi} \nabla_{\nu}\tilde{\phi} \left| \right|}}{\tilde{V}(\tilde{\phi})}.
\end{eqnarray}
This action represents the standard minimally coupled scalar field in affine spacetime \cite{kijowski1}. Following the same procedure made previously, one may derive the equations of motion (\ref{minimal affine field equations}).

This new feature of recasting nonminimally coupled scalar field dynamics to minimally coupled one through field redefinition, is restricted to affine gravity. As we have seen so far, for this transition, the conformal transformation is not avoidable in metric gravity.

Let us now go back to the form of the intrinsic curvature perturbation (\ref{metric intrinsic curvature perturbation}). Since the metric tensor is unique, the Hubble parameter then keeps the same form under the field reparametrisation (\ref{affine field transformation}). Thus, the intrinsic curvature perturbation is invariant under field redefinition
\begin{eqnarray}
\label{affine intrinsic curvature perturbation in metric gravity}
\tilde{\mathcal{R}} \equiv \frac{H}{\dot{\tilde{\phi}}} \delta \tilde{\phi}
= \frac{H}{\dot{\phi}} \delta \phi 
\equiv \mathcal{R}.
\end{eqnarray}
Unlike the metric theory case (\ref{transformed metric intrinsic curvature perturbation}), this invariant quantity would provide an invariant spectrum of density perturbation, and then a unique spectral index $n_{s}$. The same conclusion for the number of $e$-foldings $N$.

\subsection{Multifield dynamics}
\label{subsec:multifields}
Coupling matter to affine gravity is not restricted to single scalar fields, in fact, affine spacetime accommodates multifields too. The general affine action which describes the scalar fields $\phi^{A}$ coupled to the affine connection, is written as
\begin{eqnarray}
\label{affine multi nonminimal}
S[\Gamma,\phi^{A}] = \int d^{4}x \frac{\sqrt{ \left| \right| \mathcal{F}(\phi^{1},\dots,\phi^{N})
R_{\mu\nu} \left(\Gamma\right) -\delta_{AB} \nabla_{\mu}\phi^{A} \nabla_{\nu}\phi^{B} \left| \right|}}{V(\phi^{1},\dots,\phi^{N})}.
\end{eqnarray}
This action generalizes the affine theory of a single field (\ref{affine single}) and the dynamics of the fields may easily be obtained by following the same procedure made so far. The theory is valid for general nonzero potentials $V(\phi^{1},\dots,\phi^{N})\neq 0$, where one may impose some specific symmetries on the field space, like $SO(N)$ symmetry. In this particular cases, one may have to add an additional piece to the potentials to prevent the action from going singular at the poles of the potential function. This additional term may be simply a cosmological constant.     

The gravitational equations are derived by varying the last action with respect to the affine connection $\Gamma$. This leads to the following dynamical equation 
\begin{eqnarray}
\label{dynamical equation multifields}
\nabla_{\alpha} \left\lbrace \mathcal{F}(\phi^{1},\dots,\phi^{N})
\frac{\sqrt{ \left| \right|K(\Gamma,\phi^{A})  \left| \right|}}{V(\phi^{1},\dots,\phi^{N})} (K^{-1}(\Gamma,\phi^{A}))^{\mu\nu}   \right\rbrace =0,
\end{eqnarray}
where we have used for brevity the following tensor
\begin{eqnarray}
K_{\mu\nu}(\Gamma,\phi^{A})=
\mathcal{F}(\phi^{1},\dots,\phi^{N})
R_{\mu\nu} \left(\Gamma\right) -\delta_{AB} \nabla_{\mu}\phi^{A} \nabla_{\nu}\phi^{B}.
\end{eqnarray}
Solution to the dynamical equation (\ref{dynamical equation multifields}) requires an invertible tensor $g_{\mu\nu}$ where the connection is compatible with it, i.e, 
\begin{eqnarray}
\nabla_{\alpha} g_{\mu\nu}=0,
\end{eqnarray}
and satisfies the identity
\begin{eqnarray}
\sqrt{\left| \right| g \left| \right| }(g^{-1})^{\mu\nu}=
\mathcal{F}(\phi^{1},\dots,\phi^{N})
\frac{\sqrt{ \left| \right|K(\Gamma,\phi^{A})  \left| \right|}}{V(\phi^{1},\dots,\phi^{N})} (K^{-1}(\Gamma,\phi^{A}))^{\mu\nu}.
\end{eqnarray}
The last identity is nothing but a compact form of a gravitational field equations with matter and it is easy to put it in a tensor form as
\begin{eqnarray}
\mathcal{F}(\phi^{1},\dots,\phi^{N})
R_{\mu\nu} \left(\Gamma\right) -\delta_{AB} \nabla_{\mu}\phi^{A} \nabla_{\nu}\phi^{B}=g_{\mu\nu}\frac{V(\phi^{1},\dots,\phi^{N})}{\mathcal{F}(\phi^{1},\dots,\phi^{N})}.
\end{eqnarray}
Now the tensor $g_{\mu\nu}$ plays the role of a metric, and the connection $\Gamma$ is reduced to the Levi-Civita connection of this metric. This tensor can be used then for raising, lowering as well as contractions. To that end, one may write the last equation in terms of Einstein tensor as
\begin{eqnarray}
\label{affine multi gravitational equations}
\mathcal{F}(\phi^{1},\dots,\phi^{N})G_{\mu\nu}(g)=&&
\delta_{AB} \nabla_{\mu}\phi^{A} \nabla_{\nu}\phi^{B}
-\frac{1}{2}g^{\alpha\beta}\delta_{AB} \nabla_{\alpha}\phi^{A} \nabla_{\beta}\phi^{B}g_{\mu\nu}  \nonumber \\
&&-\frac{V(\phi^{1},\dots,\phi^{N})}{\mathcal{F}(\phi^{1},\dots,\phi^{N})}.
\end{eqnarray}
The equation of motion of a scalar field $\phi^{A}$ is obtained by varying with respect to $\phi^{B}$. This leads after simplification to the following equation
\begin{eqnarray}
\label{affine multi equations of motion}
\Box \phi^{A}-V_{,A}+\frac{1}{2}\mathcal{F}_{,A}R(g)+\Psi=0,
\end{eqnarray}
where the Comma refers to the derivative with respect to the field $\phi^{A}$, and the function $\Psi$ is given by
\begin{eqnarray}
\Psi=(1-\mathcal{F}^{-1})V_{,A}-\mathcal{F}^{-1}\mathcal{F}_{,A}g^{\alpha \beta}
\delta_{CD}\nabla_{\alpha}\phi^{C}\nabla_{\beta}\phi^{D}.
\end{eqnarray}
The action (\ref{affine multi nonminimal}) that leads to the complicated equations of motion (\ref{affine multi gravitational equations}) and (\ref{affine multi equations of motion}) can be recast to a simpler action which describes a minimally coupled multifields. This is done without altering the geometric part (connection or curvature), but only by a field redefinition of the form  
\begin{eqnarray}
\label{affine multifield redefinition}
d\phi^{A} \rightarrow
d \tilde{\phi}^{A}=\frac{M_{Pl}}{\sqrt{\mathcal{F}}} d\phi^{A}.
\end{eqnarray}
This reparametrisation must be followed by a potential rescaling as 
\begin{eqnarray}
\label{affine multifield potential rescaling}
V \rightarrow \tilde{V}=\frac{M^{4}_{Pl}}{\mathcal{F}^{2}} V(\phi^{1},\dots,\phi^{N}).
\end{eqnarray}
In this case, the action (\ref{affine multi nonminimal}) takes the following form
\begin{eqnarray}
\label{affine multi minimal}
S[\Gamma,\phi^{A}] \rightarrow \int d^{4}x \frac{\sqrt{ \left| \right| 
M^{2}_{Pl} R_{\mu\nu} \left(\Gamma\right) -\delta_{AB} \nabla_{\mu}\tilde{\phi}^{A} \nabla_{\nu}\tilde{\phi}^{B} \left| \right|}}{\tilde{V}(\tilde{\phi}^{1},\dots,\tilde{\phi}^{N})}.
\end{eqnarray}
This action represents the theory of multifields minimally coupled to gravity through affine connection. As can be easily checked by using the transformations (\ref{affine multifield redefinition}) and (\ref{affine multifield potential rescaling}), the gravitational equations (\ref{affine multi gravitational equations}) are reduced to the standard Einstein equations sourced by scalar fields $\tilde{\phi}^{A}$ and the same spacetime metric tensor $g_{\mu\nu}$. This is also the result one can obtain when performing the variation of action (\ref{affine multi minimal}) with respect to the connection and solve the obtained dynamical equations. This remarkable result is restricted to affine gravity where metrical properties are not defined \textit{a priori}, and then no conformal transformation makes sense. The absence of this latter prevents the appearance of the additional unwanted terms which are proportional to the field derivatives, and then provides us with a canonical kinetic terms of the fields. Different matter fields here which can be obtained from each other through field redefinition couple to the same and unique spacetime metric.   

\subsection{Vacuum energy sets metrical geometry: uniqueness of the generated frame}

Up to now, the transition between non-minimal and minimal coupling in affine gravity is shown without referring to any physical principle that underlies the equivalence of the theories. However, affine gravity based on the structure of the actions proposed so far, provides a good reason for that. 

The key point is that the affine actions are singular at $V(\phi) = 0$, which means that the scalar field must always have a non-zero potential energy. This property holds for multifields too. The nonzero potential of different fields may be described by a nonzero primordial part $V_{0}$ which keeps the affine action non-singular even in the absence of the fields. This turns out to be the vacuum energy. The presence of this quantity in the affine spacetime imposes (covariantly) an energy momentum tensor of vacuum $T_{\mu\nu}$ with a non-singular inverse $(T^{-1})^{\lambda \rho}$. This naturally defines a Levi-Civita connection as \cite{demir stress energy of vacuum}
\begin{eqnarray}
\label{GammaT}
{}^{T}\Gamma^{\lambda}_{\,\,\,\mu \nu} = \frac{1}{2} (T^{-1})^{\lambda \rho} \left(\partial_{\mu} T_{\nu \rho} + \partial_{\nu} T_{\rho \mu} - \partial_{\rho} T_{\mu\nu}\right)
\end{eqnarray}
with respect to which
\begin{eqnarray}
\label{compatibility}
\nabla^{T}_{\mu} T_{\alpha\beta} = 0.
\end{eqnarray}

Originally, it is this fundamental structure which provides a solution to the dynamical equations (\ref{dynamical equation single field}) and (\ref{dynamical equation multifields}). In fact, equation (\ref{dynamical equation single field}) is solved and put in the following form \cite{affine inflation}
\begin{eqnarray}
\label{affine single gravitational equations2}
(M_{Pl}^2 + \xi \phi^2) R_{\mu\nu} - \nabla_{\mu}\phi \nabla_{\nu} \phi=
\left(\frac{M_{Pl}^2}{M_{Pl}^2 + \xi \phi^2}\right)\frac{V(\phi)}{V(\phi_{min})}
T_{\mu\nu}.
\end{eqnarray}
The vacuum energy momentum tensor which is inherently contained in affine spacetime can be incorporated in its mixed form in terms of $V(\phi_{min})$ as
\begin{eqnarray}
\label{vacuum tensor}
T^{\mu}_{\nu} && \equiv V(\phi_{min}) \delta^{\mu}_{\nu} \\
&&= V(\phi_{min})  T_{\nu\alpha} (T^{-1})^{\alpha\mu}. \nonumber
\end{eqnarray}
The transition to minimal coupling is made by transforming the equations of motion (\ref{affine single gravitational equations2}) under the field redefinition (\ref{affine field transformation}). Since both vacuum energy $V(\phi_{min})$ and its energy momentum tensor $T^{\mu}_{\nu}$ are redefined, they form an invariant ratio
\begin{eqnarray}
\label{ration delta}
\frac{T^{\mu}_{\nu}}{V(\phi_{min})}=
\frac{\tilde{T}^{\mu}_{\nu}}{\tilde{V}[\tilde{\phi}(\phi_{min})]}
\equiv \delta^{\mu}_{\nu}.
\end{eqnarray}
This identity tensor which facilitates the covariant description of vacuum energy in affine spacetime reflects the metrical properties implicitly. In fact, the dimentionless metric tensor is nothing but the \enquote{unique} ratio
\begin{eqnarray}
\label{ration delta}
\frac{T_{\mu\nu}}{V(\phi_{min})}=
\frac{\tilde{T}_{\mu\nu}}{\tilde{V}[\tilde{\phi}(\phi_{min})]}
\equiv g_{\mu\nu}.
\end{eqnarray}
With this metric tensor at hand, the gravitational equations can be recast to a minimally coupled case without conformal transformation. Figure \ref{fig:rectangular shap} above shows the rescaling of vacuum energy momentum tensor when performing a field redefinition, and the invariant (unique) metric tensor.

\begin{figure}[t]
\begin{center} 
\begin{tikzpicture} 
\matrix(cd)[matrix of math nodes,
  row sep=4.6em, column sep=3.5cm, 
  text height=1.5ex, text depth=0.50ex]{
  V(\phi) & \tilde{V}[\tilde{\phi}(\phi)] \\ 
  V(\phi_{\min}) & \tilde{V}[\tilde{\phi}(\phi_{min})] \\ 
  T^{\mu}_{\nu}=V(\phi_{min}) \delta^{\mu}_{\nu} & \tilde{T}^{\mu}_{\nu}=\tilde{V}[\tilde{\phi}(\phi_{min})] \delta^{\mu}_{\nu} \\
}; 
\draw[->] (cd-1-1) edge node[label=above:$(\ref{affine field transformation})$] (U) {} (cd-1-2); 
\draw[->] (cd-2-1) edge node[label=above:$(\ref{affine field transformation})$] (V) {} (cd-2-2); 
\draw[->] (cd-3-1) edge node[label=above:Vacuum rescaling] (W) {} (cd-3-2); 
\draw[->] (cd-1-1) edge node[label=left: ] (U) {} (cd-2-1); 
\draw[->] (cd-2-1) edge node[label=right:$(\ref{vacuum tensor})$] (V) {} (cd-3-1); 
\draw[->] (cd-1-2) edge node[label=right: ] (W) {} (cd-2-2); 
\draw[->] (cd-2-2) edge node[label=left:$(\ref{vacuum tensor})$] (U) {} (cd-3-2); 
\end{tikzpicture} 
\caption{The transition to minimal coupling in affine gravity is performed through field redefinition. The transformations of vacuum energy and its associated energy momentum tensor provide an invariant and unique ratio that represents a dimensionless metric tensor.}
\label{fig:rectangular shap}
\end{center}
\end{figure}

\section{Induced gravity: metric or affine structure?}
\label{sec: affine gravity induces metric gravity}

As we have seen so far, the problems of frames have their origin from the \enquote{metric} conformal transformation (\ref{conformal transformation}). This transformation becomes necessary only for generalized theories of gravity such as the one given by action (\ref{non-minimal coupling metric action}). In addition to Ricci scalar, this type of theories may include higher order curvature terms which can be generated automatically through field quantization in curved spacetime. This standard view of induced gravity requires a metrical structure which is necessary in the framework of field theory in curved background. However, as we shall see in this section, this structure may not be postulated \textit{a priori}, but gravity as the metric elasticity of space will be induced from affine connection and scalar fields \cite{induced affine inflation}. 

\subsection{Induced \enquote{metric} gravity assumes metric structure \textit{a priori}}

It has been known for a long time that nonminimal couplings like the last term of action (\ref{non-minimal coupling metric action}) arise in most of the theories in curved spacetime, and are generated from quantum corrections to matter by integrating high energy modes. It turned out that this procedure may give a quantum origin to Newton's constant without referring to it classically in the initial action (\ref{non-minimal coupling metric action}).

In the context of induced gravity, classical scalar fields may live in a curved background described by a Lorentzian manifold. In this view, although metric properties of spacetime take place in the manifold, gravity which is described by Einstein-Hilbert action is absent. This setup is usually put in a standard form as
\begin{eqnarray}
\label{induced gravity action-metric}
S=\int d^{4}x \sqrt{-g}\left[\frac{1}{2}\xi \phi^{2}R\left(g\right)-\frac{1}{2}
g^{\mu\nu} \nabla_{\mu}\phi\nabla_{\nu}\phi -V(\phi)  \right].
\end{eqnarray}

There are two ways of inducing gravity in this setup. The first and straightforward case is that Newton's constant (and then Einstein-Hilbert action) arises at a classical constant background of the field $\phi$. This can easily be seen when $\phi=v$, where the first term of (\ref{induced gravity action-metric}) is reduced to Einstein-Hilbert action with Newton's constant \cite{zee}
\begin{eqnarray}
G_{N}=(8\pi \xi v^{2})^{-1}.
\end{eqnarray}

In Sakharov's view and its modern perspectives, gravity is induced from the contribution of the one loop  effective action of (\ref{induced gravity action-metric}). This contribution reads
\begin{eqnarray}
\label{action00}
\Delta S=-\frac{1}{2} \text{Tr} \left\lbrace \ln \left[\Box +V^{\prime \prime} +\xi R\left(g\right) \right] \right\rbrace,
\end{eqnarray}
where $V^{\prime \prime}$ is the second derivative of the potential evaluated at the background field.

By adopting an explicit UV cutoff ($\Lambda_{UV}$) and regularising the action, curvature terms including the Ricci scalar appear automatically. The generation of these terms is followed by quantum corrections to the potential. In this process, Newton's constant is induced at one loop as \cite{sakharov}
\begin{eqnarray}
G_{N} \sim 1/\Lambda_{UV}^{2},
\end{eqnarray}
leading to the associated experimental value at the Planck scale $\Lambda_{UV}=M_{Pl}$.

This modern view of gravity as an induced phenomenon rather than a fundamental force got much attention, since it provides a possible connection between particle physics and gravity \cite{sakharov,naturalness}.

Although gravity is induced here by generating Einstein-Hilbert action, the setup may lack the concrete emergence of this force in terms of the metrical properties of spacetime. In fact, a key element in Einstein's general theory of relativity is the metric tensor which is postulated \textit{a priori} in the Lorentzian spacetime manifold. Like general relativity, this metric structure is already assumed in induced gravity.

\subsection{Affine gravity as an origin of metric elasticity of space}
\label{sec:Affine gravity as an origin of metric elasticity of space}
Here we will address a possible fundamental origin of metric gravity itself. Our setup will be based on the following assumptions:
\begin{enumerate}
\item Spacetime is affine, i.e, it is endowed with an affine connection $\Gamma$ that makes comparison of vector and tensors at different points possible through parallel displacement, without referring to distances and angle measurements.
\item This geometry accommodates scalar fields $\phi$ with non-vanishing potential $V(\phi)$ ensuring non-vanishing vacuum energy.
\end{enumerate}
It is clear from these properties that our spacetime geometry does not recognize the metric structure. In other word, gravity \`{a} la Einstein is completely absent.

To that end, this primary theory is described by the following action
\begin{eqnarray}
\label{primary action}
S \left[\Gamma, \phi \right]= \int d^{4}x \frac{\sqrt{\left| \right| \xi \phi^2 R_{\mu\nu}\left(\Gamma\right)- \nabla_{\mu}\phi\nabla_{\nu}\phi \left| \right|}}{V(\phi)},
\end{eqnarray}
where $R_{\mu\nu}(\Gamma)$ is the Ricci tensor constructed from the affine connection $\Gamma$ and $\xi$ is a constant.

Our guiding principle in writing action (\ref{primary action}) is the induction of gravity through two important steps \cite{induced affine inflation}:
\begin{enumerate}
\item First, by inducing the scale of gravity, in the philosophy of action (\ref{transformed affine action}), from vacuum expectation value of heavy scalars via spontaneous symmetry breaking.
\item Emergence of the metric tensor from nonzero vacuum energy.
\end{enumerate}
The first step is realized when the potential attains its minimum at some energy scale $v$ where
\begin{eqnarray}
\label{ssb potential}
V(\phi)=V_{0}+\frac{\lambda}{4}(\phi^{2}-v^{2})^{2},
\end{eqnarray}
and then the nonminimal coupling term in (\ref{primary action}) acquires a vacuum expectation value
\begin{eqnarray}
\xi v^{2} R_{\mu\nu}(\Gamma).
\end{eqnarray}
The fundamental scalar of gravity $M^{2}_{Pl}$ arises then in pure affine spacetime as 
\begin{eqnarray}
M^{2}_{Pl}=\xi v^{2},
\end{eqnarray}
where the constants $\xi$ and $v$ must ensure the value $M_{Pl}\simeq
2.4\times 10^{18}$.

At the vacuum, $\phi=v$, the potential is left only with a vacuum energy $V_{0}$. This piece must not vanish since it protects the affine action (\ref{primary action}) from going singular. This explains its necessity in (\ref{ssb potential}), unlike in GR where its absence has no effects. 

This nonzero vacuum energy is the fundamental quantity behind the emergence of the metric in our second step.

The dynamical equations arising from variation of action (\ref{primary action}) with respect to $\Gamma$ take the form
\begin{eqnarray}
\label{dynamical equation}
\nabla_{\mu} \left\lbrace \xi \phi^{2} \frac{ \sqrt{|| K(\Gamma,\phi)||}}{V(\phi)}
\left( K^{-1} \right)^{\alpha \beta} \right\rbrace = 0, 
\end{eqnarray}
where again for simplicity we have put
\begin{eqnarray}
\label{tensor k}
K_{\mu\nu}(\Gamma,\phi)=
\xi \phi^{2} R_{\mu\nu}(\Gamma)-\nabla_{\mu}\phi \nabla_{\nu}\phi.
\end{eqnarray}
Solution to this equation is given in terms of a rank-two tensor $g_{\mu\nu}$, such that
\begin{eqnarray}
\label{density equality}
M^{2}\sqrt{||g||}(g^{-1})^{\mu\nu}=
\xi \phi^{2} \frac{ \sqrt{|| K(\Gamma,\phi)||}}{V(\phi)}
\left( K^{-1} \right)^{\mu \nu}
\end{eqnarray}
and
\begin{eqnarray}
\label{compatibility condition}
\nabla_{\alpha} g_{\mu\nu}=0,
\end{eqnarray}
where $M$ is a mass constant.

By the same argument made in the last section, now the affine connection is reduced to the metric connection of the emerged tensor $g_{\mu\nu}$ which plays the role of the metric tensor. 

It is important to notice the case $\left\langle \phi \right\rangle=v$, where the metric tensor (\ref{density equality}) becomes finite only for $V(v)\neq 0$. This nonzero vacuum energy guarantees the emergence of the metric.

For $\phi=v$, the gravitational equations (\ref{density equality}) are equivalent to Einstein's equations with a cosmological constant, and this leads to 
\begin{eqnarray}
\label{planck mass}
M^{2}=\xi v^{2}= M^{2}_{Pl}.
\end{eqnarray}
In general, however, the theory described by action (\ref{primary action}) is not equivalent to metric induced gravity (\ref{induced gravity action-metric}), and the resulting field equations (\ref{density equality}) can be written in a standard form as   
\begin{eqnarray}
\label{induced affine gravitational equations}
\xi \phi^{2} G_{\mu\nu}=\nabla_{\mu}\phi \nabla_{\nu}\phi
-\frac{1}{2}g_{\mu\nu}\nabla^{\lambda}\phi \nabla_{\lambda}\phi -g_{\mu\nu}V(\phi)\left( \frac{M^{2}}{\xi \phi^{2}}\right).  
\end{eqnarray}
Now variation with respect to the field $\phi$ leads to the following equation of motion \cite{induced affine inflation}
\begin{eqnarray}
\label{equ of motion of phi}
\Box \phi-V^{\prime}\left(\phi\right)+\xi\phi R\left(g\right)+\Psi\left(\phi\right)=0,
\end{eqnarray}
where the function $\Psi$ is given by
\begin{eqnarray}
\label{Psi}
\Psi\left(\phi\right)=\left(1 -\frac{M^{2}}{\xi \phi^{2}} \right)V^{\prime}\left(\phi\right)-\frac{2}{\phi}\left(\nabla\phi \right)^{2}.
\end{eqnarray}
With these field equations and the compatibility condition (\ref{compatibility condition}) which appear \textit{a posteriori}, the affine theory is reduced to metric theory. Metric elasticity of space becomes an emergent phenomenon where the concept of distances and angles arise only at a final stage. This stage is represented by the large scale structure of spacetime. It is for no reason that the latter could have started with the familiar metric structure at very early times. In fact, the existence of singular regions in space, such as black holes and the initial singularity (big-bang) suggest a completely different structure for spacetime. In these small regions of space where quantum effects, translated by Heisenberg uncertainty principle, are not avoidable, distances and clock rates measurements break down \cite{murphy}. These concepts at large scales may have arisen from a simpler structure of spacetime, which is endowed with an affine connection and a nonzero vacuum energy given in terms of   
\begin{eqnarray}
V_{0} \sim m_{\nu}^{4},
\end{eqnarray}
where $m_{\nu}$ is the Neutrino mass.

This induced affine gravity is able to give an origin to not only the scale of gravity as in \cite{zee} but also the metrical structure. It could be also interesting if one accomplishes this via the loops of matter fields \cite{azri-future work}.

The affine approach to gravity which we have discovered in this short review stands viable framework to study scalar field dynamics and then it must reveal interesting results when applied to cosmology and astroparticle physics. As we have shown throughout this review, affine gravity acquires a unique description in a sense that it prefers only one metric tensor for different couplings. Nonminimally coupled field dynamics can be transformed into minimally coupled ones with a modified
potential but the same metric tensor. Thus, there is no mixing between \enquote{geometry} and scalar fields in the transition process. This new feature, which is not valid in metric theories, plays an important role in avoiding the ambiguities of conformal frames that arise in cosmological inflation \cite{affine inflation, induced affine inflation}. Although the problem is somehow settled at the classical level, and the frames can be considered equivalent in metric theories, the ambiguity arises when treating the quantum fluctuations of the inflaton. In fact, since the new metric tensor (\ref{conformal transformation}) includes the scalar degree of freedom, the latter is automatically mixed with tensor modes. This fact would lead to difficulties of obtaining the same results, when the physical quantities are transformed back to the original frame. 

The inflaton fluctuations enter the definition of an important quantity; the intrinsic curvature perturbation, which is the basis of the \textit{slow-roll approximation} underlying the inflationary regime. The non (conformal) invariance of the perturbation (\ref{transformed metric intrinsic curvature perturbation}) leads clearly to different predictions in different frames. In affine gravity however, the metric tensor is unique, the calculation in affine gravity is protected from the mixing of scalar and tensor degrees of freedom that arise from transformations like (\ref{conformal transformation}). This also can be translated by the invariance of the intrinsic curvature perturbation which is the basis of the perturbation calculations.

In the next section, we will summarize the inflationary dynamics of different models in the context of affine gravity.

\section{Affine inflationary models}
\label{sec:inflationary models}

\subsection{$\phi^{4}$-affine inflation}

Standard affine inflation is the inflationary phase which is based on the affine gravity action (\ref{affine single}). Here, different type of potentials leads to different affine inflationary models. In what follows, we will study affine inflation driven by the following simple potential
\begin{eqnarray}
V(\phi)=\frac{\lambda_{\phi}}{4}\phi^{4}.
\end{eqnarray}
This potential has been studied in details in metric gravity, and for this reason we have proposed it here in order to show the differences between the two theories.  

To simplify the calculation we will apply the field and potential redefinitions (\ref{affine field transformation}). In this case, we have shown that the equations of motion take the standard forms (\ref{minimal-affine-grav-eq}) and (\ref{minimal affine field equations}). One may easily integrate equation (\ref{affine field transformation}) and get
\begin{eqnarray}
\phi(\tilde{\phi})=\frac{M_{Pl}}{\sqrt{\xi}}\sinh\left(\frac{\sqrt{\xi}}{M_{Pl}}\tilde{\phi}\right).
\end{eqnarray}
In this case the potential in (\ref{affine field transformation}) takes the form
\begin{eqnarray}
\tilde{V}(\tilde{\phi})=\frac{\lambda_{\phi}}{4}\frac{M^{4}_{Pl}\xi^{-2}\sinh^{4}\left(\frac{\sqrt{\xi}}{M_{Pl}}\tilde{\phi}\right)}
{\left(1+\sinh^{2}\left(\frac{\sqrt{\xi}}{M_{Pl}}\tilde{\phi}\right)\right)^{2}}.
\end{eqnarray}
Now, for large fields $\tilde{\phi} >M_{Pl}/\sqrt{\xi}$, we easily calculate the slow roll parameters as
\begin{eqnarray}
\label{slow roll parameters}
\epsilon=\frac{M^{2}_{Pl}}{2}\left(\frac{\tilde{V}^{\prime}}{\tilde{V}}  \right)^{2}
\simeq 128\xi \exp \left(-4\frac{\sqrt{\xi}}{M_{Pl}}\tilde{\phi} \right) \\
\eta= M^{2}_{Pl} \left(\frac{\tilde{V}^{\prime\prime}}{\tilde{V}}  \right)
\simeq -32\xi \exp \left(-2\frac{\sqrt{\xi}}{M_{Pl}}\tilde{\phi} \right) \\
\zeta^{2}= M^{4}_{Pl}\frac{\tilde{V}^{\prime\prime\prime}\tilde{V}^{\prime}}{\tilde{V}^{2}} \simeq
\left(32 \xi\right)^{2}\exp \left(-4\frac{\sqrt{\xi}}{M_{Pl}}\tilde{\phi} \right).
\end{eqnarray}
The same for the number of e-foldings which reads
\begin{align}
N&=\frac{1}{M^{2}_{Pl}}\int_{\tilde{\phi}_{f}}^{\tilde{\phi}_{i}}\frac{\tilde{V}(\tilde{\phi})}{\tilde{V}^{\prime}(\tilde{\phi})}d\tilde{\phi} \nonumber \\ &
\simeq\frac{1}{32\xi}\left[\exp\left(2 \frac{\sqrt{\xi}}{M_{Pl}}\tilde{\phi}_{i}\right)- 
\exp\left(2 \frac{\sqrt{\xi}}{M_{Pl}}\tilde{\phi}_{f}\right)\right].& \label{efoldings1}
\end{align}
The final value $\tilde{\phi}_{f}$ is obtained from $\epsilon=1$ when inflation ends. In this case we obtain
\begin{eqnarray}
\frac{\tilde{\phi}_{f}}{M_{Pl}}=\frac{\ln(128\xi)}{4\sqrt{\xi}}.
\end{eqnarray}
The initial value is obtained in terms of the number of e-foldings. From equation (\ref{efoldings1}) one may easily find 
\begin{eqnarray}
\label{initial fi}
\frac{\tilde{\phi}_{i}}{M_{Pl}}=\frac{\ln (32\xi N)}{2\sqrt{\xi}}.
\end{eqnarray}
Finally, using the previous parameters, the spectral index at first order, $n_{s}=1-6\epsilon+2\eta$, reads
\begin{eqnarray}
n_{s}\simeq 1-\frac{3}{4\xi N^{2}}-\frac{2}{N}.
\end{eqnarray}
This is clearly completely different than that of metric gravity \cite{kaiser2}
\begin{eqnarray}
n_{s}\simeq 1-\frac{32\xi}{16\xi N-1}.
\end{eqnarray}
The affine inflation tensor-to-scalar ratio reads
\begin{eqnarray}
r= 16\epsilon \simeq \frac{2}{\xi N^{2}},
\end{eqnarray}
where we have used solution (\ref{initial fi}).
 
Observational bounds on the spectral index imply that the nonminimal coupling parameter must satisfy $\xi \gtrsim 3.12 \times 10^{-2} $. For $60$ e-foldings, 
the ratio $r$ has an upper bound
\begin{eqnarray}
r \lesssim 1.7 \times 10^{-2},
\end{eqnarray}
Thus, affine inflation predicts a small amount of tensor perturbations which is in the range of the observed value \cite{planck}. A large $\xi$ however produces a negligible ratio.
\subsection{Induced inflation: Illustrative example of frame ambiguities}
\label{higgs inflation}

\subsubsection{Induced affine inflation}

Induced affine inflation is the inflationary dynamics based on action (\ref{primary action}) of section \ref{sec:Affine gravity as an origin of metric elasticity of space}. The standard induced gravity potential is given as follows
\begin{eqnarray}
\label{inflationary potential}
V(\phi)= V_{0}+\frac{\lambda}{4}(\phi^{2}-v^{2})^{2},
\end{eqnarray}
Below, we assume that the universe is described by the FRW metric with the scale factor $a\left(t\right)$. Then cosmological dynamics of the inflaton $\phi\left(\vec{\text{x}},t\right)$ is described by 
\begin{eqnarray}
\ddot{\phi}+3H\dot{\phi}-\frac{\dot{\phi}^{2}}{\phi}+\frac{(\vec{\nabla}\phi)^{2}}
{a^{2}\phi}-\frac{\vec{\nabla}^{2}\phi }{a^{2}} 
=\frac{4M^{2}}{\xi \phi^{3}}V\left(\phi \right)
-\frac{M^{2}}{\xi \phi^{2}}V^{\prime}\left(\phi\right), 
\end{eqnarray}
where
\begin{eqnarray}
H^{2}=\frac{1}{3\xi \phi^{2}}\left(\frac{\dot{\phi}^{2}}{2} +\frac{M^{2}}{\xi \phi^{2}}V\left(\phi\right) \right)
\end{eqnarray}
is the Hubble parameter. 

Here we have used the gravitational equations (\ref{induced affine gravitational equations}) and the field equation of the scalar field (\ref{equ of motion of phi}).

Inflation proceeds slowly if the slow-roll conditions
\begin{eqnarray}
\frac{\dot{\phi}}{\phi}\ll H,\,\,\,\,\text{and},\,\,\,
\dot{\phi}^{2} \ll \frac{M^{2}}{\xi \phi^{2}}V\left(\phi\right)
\end{eqnarray}
are satisfied. 

Under these conditions, the background field evolves as
\begin{eqnarray}
\label{slow roll equation1}
3H\dot{\phi}
\simeq \frac{4M^{2}}{\xi \phi^{3}}V\left(\phi \right)
-\frac{M^{2}}{\xi \phi^{2}}V^{\prime}\left(\phi\right),
\end{eqnarray}
\begin{eqnarray}
H^{2}\simeq \frac{M^{2}}{3\xi^{2}\phi^{4}} V\left(\phi\right),
\label{slow roll equation2}
\end{eqnarray}
To solve for the background field $\phi(t)$, we write equation (\ref{slow roll equation1}) as
\begin{eqnarray}
\frac{d\phi}{dt}=4\xi \phi \left[1-\frac{1}{\left(1-\frac{v^{2}}{\phi^{2}} \right)} \right]H,
\end{eqnarray}
where we have used the potential (\ref{inflationary potential}).

Now, using the Hubble parameter (\ref{slow roll equation2}), we easily get
\begin{eqnarray}
\frac{d\phi}{dt}=
\pm \frac{2Mv^{2}}{\phi}\sqrt{\frac{\lambda}{3}},
\end{eqnarray}
which can be integrated easily as
\begin{eqnarray}
\label{solution phi}
\phi^{2}\left(t\right)=\phi_{i}^{2}
\pm 4Mv^{2}\sqrt{\frac{\lambda}{3}}\,t,
\end{eqnarray}
where \enquote{i} denotes the initial values.

Now let us turn to the scale factor $a(t)$. This can be obtained by firstly dividing both sides of equation (\ref{slow roll equation1}) by $H^{2}$, then
\begin{eqnarray}
\frac{\dot{\phi}}{H}=
\frac{4M^{2}}{3\xi \phi^{3}}\frac{V(\phi)}{H^{2}}
-\frac{M^{2}}{3\xi\phi^{2}}\frac{V^{\prime}(\phi)}{H^{2}}
\end{eqnarray}
Using equation (\ref{slow roll equation2}) for $H^{2}$ and (\ref{inflationary potential}) for $V(\phi)$, we get
\begin{eqnarray}
\frac{\dot{\phi}}{H}=
4\xi \phi \left[1-\frac{1}{\left(1-\frac{v^{2}}{\phi^{2}} \right)} \right]
\end{eqnarray}
or
\begin{eqnarray}
\frac{da(t)}{a(t)}=
\frac{d\phi}{4\xi \phi \left[1-\frac{1}{\left(1-\frac{v^{2}}{\phi^{2}} \right)} \right]}.
\end{eqnarray}
This is easily integrated as
\begin{eqnarray}
\frac{a\left(t\right)}{a_{i}}=\left(\frac{\phi\left(t\right)}{\phi_{i}} \right)
^{1/4\xi}\exp\left\lbrace \frac{1}{8\xi v^{2}}\left(\phi_{i}^{2}-\phi^{2}\left(t\right) \right)  \right\rbrace,
\label{solution a}
\end{eqnarray}
where $\phi_{i}$ and $a_{i}$ are the initial values.

In standard induced gravity inflation, we are interested in small fields where $\phi << v$. In this case, the last equation leads to
\begin{eqnarray}
\label{scale factor power low}
a\left(t\right) \propto t^{1/8\xi}.
\end{eqnarray}
To calculate the scalar spectral index of this model, we first write down the equations of motion of the quantum fluctuations. Then, expanding $\phi\left(\vec{\text{x}},t\right)$ as $\phi\left(\vec{\text{x}},t\right)=\phi \left(t\right)+\delta\phi\left(\vec{\text{x}},t\right)$ where the background field $\phi \left(t\right)$ is given by (\ref{solution phi}), it is easy to see that the fluctuations obey the equation
\begin{eqnarray}
\label{fluctuation equation}
\ddot{\delta\phi}+3H\dot{\delta\phi}+\frac{k^{2}}{a^{2}}\delta\phi
\simeq \frac{\lambda M^{2}v^{2}}{\xi \phi^{2}}\left(1-\frac{3v^{2}}{\phi^{2}} \right)
\delta\phi,
\end{eqnarray}
where $\vec{k}$ is the momentum component corresponding to $\vec{x}$.

Here we will be interested in the case where the term $k^{2}/a^{2}$ dominates the term on the right-hand side at the time of the last horizon crossing. This simplifies the last equation which is now approximated to the equation of a massless field. 

We proceed in a standard way by using the conformal time $d\eta=dt/a(t)$, and the conformal field $\psi=a\delta\phi$. In this case, the equation of motion (\ref{fluctuation equation}) reads 
\begin{eqnarray}
\psi^{\prime \prime}_{k}
-\frac{2(1-4\xi)}{(1-8\xi)^{2}\eta^{2}}\psi_{k}+k^{2}\psi_{k} \simeq 0,
\end{eqnarray}
where we have used the scale factor (\ref{scale factor power low}) which gives
\begin{eqnarray}
\label{hubble in terms of eta}
\frac{a^{\prime \prime}}{a}=\frac{2(1-4\xi)}{(1-8\xi)^{2}\eta^{2}},
\end{eqnarray}
and prime is the derivative with respect to the conformal time.

Our aim is to put the last equation in a Bessel's equation form. For this, we define the function $v$ as $v=\eta^{-1/2}\psi$ and use the notation $x=k\eta$, then we get
\begin{eqnarray}
\frac{d^{2}v_{k}}{dx^{2}}+\frac{1}{x}\frac{dv_{k}}{dx}+
\left[1-\frac{1}{x^{2}}\frac{(3-8\xi)^{2}}{4(1-8\xi)^{2}} \right]v_{k} 
\simeq 0.
\end{eqnarray}
This a standard Bessel's equation where solutions are given in terms of Hankel function $H_{\nu}(k\eta)$ such that
\begin{eqnarray}
\label{nu}
\nu= \frac{(3-8\xi)}{2(1-8\xi)}.
\end{eqnarray}
Finally, the solution for the fluctuations $\delta \phi_{k}$ is given in a standard form as
\begin{eqnarray}
\label{eta dependence}
\delta \phi_{k} \sim \left[A_{k}H^{1}_{\nu}(k\eta)+B_{k}H^{2}_{\nu}(k\eta) \right] \eta^{\nu}.
\end{eqnarray}
Now, the important quantity is the two-point correlation function which is given by \cite{abbott,lyth}
\begin{eqnarray}
\label{correlation}
|\Delta \phi(\vec{k},\eta)|^{2}= k^{3}
\int \frac{d^{3}x}{(2\pi)^{3}}e^{i\vec{k}\vec{x}}
\left\langle \delta\phi(\vec{x},\eta)\delta\phi(0,\eta) \right\rangle
\end{eqnarray}
To calculate the scalar spectral index, we will be interested only in the $k$ dependence of the last expressions. Taking $k\eta \rightarrow 0$, the fluctuations (\ref{eta dependence}) go as $k^{-\nu}$, and then the correlation function (\ref{correlation}) goes as $|\Delta \phi(\vec{k},\eta)|^{2} \sim k^{3-2\nu}$. From this, the spectrum of density perturbation takes the form
\begin{eqnarray}
\label{spectrum of density perturbation}
\mathcal{P} \propto k^{3-2\nu},
\end{eqnarray}
which leads to the scalar spectral index $n_{s}$
\begin{eqnarray}
\label{definition of spectral index}
n_{s}-1 \equiv \frac{d\ln \mathcal{P}}{d \ln k}=3-2\nu.
\end{eqnarray}
Using equation (\ref{nu}), we easily get
\begin{eqnarray}
\label{spectral index}
n_{s}=1-\frac{16\xi}{1-8\xi}.
\end{eqnarray}

\subsubsection{Transition to minimal coupling}

Here the equations of motion are Einstein's field equations in the presence of a canonical field $\tilde{\phi}$. In this case, the field fluctuations $\delta\tilde{\phi}$ satisfy
\begin{eqnarray}
\delta \ddot{\tilde{\phi}}+3H \delta\tilde{\phi}-\frac{\vec{\nabla}^{2}\delta\phi}{a^{2}}+\tilde{V}^{\prime \prime}(\tilde{\phi})\delta\tilde{\phi}=0.
\end{eqnarray}
Interestingly, the scale factor is not altered by the field redefinition since the later does not include a metric transformation. Thus, for large fields we still have 
\begin{eqnarray}
a(t)= t^{1/8\xi}.
\end{eqnarray}
This again leads to the same term (\ref{hubble in terms of eta}) for $a^{\prime \prime}/a$ which in turn gives the same $\nu$ as in (\ref{nu})
\begin{eqnarray}
\label{nu2}
\tilde{\nu}=\nu= \frac{(3-8\xi)}{2(1-8\xi)}.
\end{eqnarray}
Finally, the fluctuations $\delta \tilde{\phi}_{k}(\eta)$ are given as 
\begin{eqnarray}
\label{eta dependence2}
\delta \tilde{\phi}_{k} \sim \left[\tilde{A}_{k}H^{1}_{\nu}(k\eta)+\tilde{B}_{k}H^{2}_{\nu}(k\eta) \right] \eta^{\nu}
\end{eqnarray}
This leads to the same $k$ dependence as (\ref{spectrum of density perturbation}) for the spectrum of perturbation
\begin{eqnarray}
\label{spectrum of density perturbation2}
\tilde{\mathcal{P}} \propto k^{3-2\tilde{\nu}},
\end{eqnarray}
from which we get a similar and unique spectral index
\begin{eqnarray}
\label{spectral index2}
\tilde{n}_{s}=1-\frac{16\xi}{1-8\xi}.
\end{eqnarray}

Let us turn now to metric induced gravity based on action (\ref{induced gravity action-metric}) and discuss briefly the associated solutions. Here, the conformal transformation which alters the form of the scale factor and then leads to different power law, would clearly provide some (though slight) difference between the density perturbations which are calculated in two conformal frames. In fact, in Jordan frame where the inflaton is described by the field $\phi$, the scale factor is given by \cite{kaiser2}
\begin{eqnarray}
a(t) \propto t^{\frac{1+6\xi}{4\xi}},
\end{eqnarray}
This leads to a power spectrum of the form
\begin{eqnarray}
\mathcal{P} \propto k^{3-2\nu},
\end{eqnarray}
where in this case \cite{kaiser2}
\begin{eqnarray}
\nu =\frac{3+14\xi}{2(1+2\xi)}.
\end{eqnarray}
Thus the spectral index is obtained from
\begin{eqnarray}
\label{definition of spectral index-jordan}
n_{s}-1 \equiv \frac{d\ln \mathcal{P}}{d \ln k}=-\frac{8\xi}{1+2\xi}.
\end{eqnarray}
Now mapping to Einstein frame affects the scale factor $a(t)$ due to the conformal transformation. In this frame we have for the fluctuations $\delta\tilde{\phi}$  
\begin{eqnarray}
\delta \ddot{\tilde{\phi}}+3\tilde{H} \delta\tilde{\phi}-\frac{\vec{\nabla}^{2}\delta\phi}{\tilde{a}^{2}}+\tilde{V}^{\prime \prime}(\tilde{\phi})\delta\tilde{\phi}=0,
\end{eqnarray}
where in this case, the scale factor and the Hubble parameter are given in terms of the scale factor and Hubble parameter of Jordan frame as in (\ref{cosmic time relation}) and (\ref{hubble relation}) respectively. In this frame we find   
\begin{eqnarray}
\tilde{a}(\tilde{t}) \propto \tilde{t}^{(1+10\xi)/8\xi}.
\end{eqnarray}
As in \cite{kaiser2}, this leads to
\begin{eqnarray}
\tilde{\nu} = \frac{3+22\xi}{2(1+2\xi)},
\end{eqnarray}
and a scalar spectral index
\begin{eqnarray}
\tilde{n}_{s}-1\equiv 3-2\nu =-\frac{16\xi}{1+2\xi}.  
\end{eqnarray}
This is clearly different from the result (\ref{definition of spectral index-jordan}) of Jordan frame. However, the differences are negligible for small $\xi$ where the predicted results are in the observed bounds.  

In \cite{induced affine inflation}, it has been shown that for induced gravity, the recent Planck bound on the tensor-to-scalar ratio $r<0.12$ implies $\xi < 10^{-3}$. This clearly drags the spectral index (\ref{spectral index}) up to its required bound. Thus, the induced gravity inflation, in both metrical and affine gravity setups, cannot satisfy the recent Planck bounds on $r$ and $n_{s}$ simultaneously. The reason is that induced gravity inflation supports only large tensor-to-scalar ratio, a feature which is not specific to induced affine gravity; it already happens in the metric induced gravity.

\subsection{Higgs affine inflation}
\label{higgs inflation}

Like any scalar field, the SM Higgs boson may drive the cosmic inflation. In this case, the predictions must be in agreement with the SM measured parameters such as the Higgs mass and the self coupling parameter. However, for a Higgs boson minimally coupled to metric gravity (GR), the observed power spectrum requires an extremely small quartic coupling $\lambda \simeq \mathcal{O}(10^{-13})$. Nevertheless, it has been shown that, this constraint can be relaxed by adding a nonminimal coupling term, Higgs-curvature, to the action. Then, the SM quartic coupling $\lambda \simeq \mathcal{O}(10^{-1})$ is attained for large nonminimal coupling parameter $\xi \simeq 10^{4}$. The nonminimal coupling then motivates the SM Higgs inflation, where the predictions are in agreement with recent Planck results \cite{higgs inflation, planck}.

Our aim here is to study \enquote{Higgs affine inflation}, where the SM Higgs is supposed to be coupled to affine gravity rather than metric gravity.

The theory is supposed to be described by the following action
\begin{eqnarray}
\label{higgs affine action}
S[\Gamma ,h] = \int d^{4}x \frac{\sqrt{ \left| \right| \left(M^2 + \xi h^2\right)R_{\mu\nu}\left(\Gamma\right) - \partial_{\mu}h \partial_{\nu}h \left| \right|}}{V(h)},
\end{eqnarray}
where we have used the unitary gauge $H=h/\sqrt{2}$ which leaves only one scalar degree of freedom with a nonzero vacuum expectation value $v$. In this case the potential is taken of the form
\begin{eqnarray}
V\left(h\right)= V_{0}+\frac{\lambda}{4}\left(h^{2}-v^{2} \right)^{2},
\end{eqnarray}
where $V_{0} \simeq m^{4}_{\nu}$ defines the observed cosmological constant $\Lambda\simeq V_{0}/M^{2}_{Pl}$, and saves action (\ref{higgs affine action}) from going singular at the vacuum $v$.

Up to now, we did not propose a unified \enquote{affine} action that incorporates all the standard model fields, this may not trivial, nevertheless, during inflation and before the reheating phase, fermions and gauge bosons maybe neglected.

Now we apply the field redefinition given in section \ref{sec: field re-parametrization} to bring the action (\ref{higgs affine action}) to a minimal coupled field $\chi$ that satisfies standard Einstein's equations. In this case, the rescaled field and the associated potential are written as
\begin{eqnarray}
\label{field and potential transformation}
\frac{d\chi}{d h}=\sqrt{1+\frac{\xi h^{2}}{M_{Pl}}} \quad, \quad \quad
U\left(\chi\right)=\frac{1}{\mathcal{F}^{2}(\chi)}
\frac{\lambda}{4}\left(h^{2}(\chi)-v^{2} \right)^{2},
\end{eqnarray}
where
\begin{eqnarray}
\mathcal{F}(h)=1+\frac{\xi h^{2}}{M^{2}_{Pl}}.
\end{eqnarray}
For small fields, i.e, $\sqrt{\xi} \, h/M_{Pl} \ll 1$, we have $\mathcal{F} \simeq 1$ and then $\chi \simeq h$, however, significant differences arise for large values of $h$ where
\begin{eqnarray}
\chi \simeq \frac{\sqrt{\xi}}{2}\frac{h^{2}}{M_{Pl}}.
\end{eqnarray}
In this case, the slow roll parameters take the following forms
\begin{eqnarray}
\label{slow roll parameters}
\epsilon=\frac{M^{2}_{Pl}}{2}\left(\frac{1}{U\left(\chi \right)} \frac{d U}{d \chi} \right)^{2}
\simeq 128\xi \exp \left(-\frac{2 \xi h^{2}}{M_{Pl}^{2}}\right) \\
\eta= M^{2}_{Pl} \left(\frac{1}{U\left(\chi \right)} \frac{d^{2} U}{d \chi^{2}} \right)
\simeq -32\xi \exp \left(-\frac{\xi h^{2}}{M_{Pl}^{2}} \right) \\
\zeta^{2}= M^{4}_{Pl}
\left(\frac{1}{U^{2}}\frac{d^{3} U}{d \chi^{3}} \frac{d U}{d \chi} \right)
\simeq
\left(32 \xi\right)^{2}\exp \left(-\frac{2\xi h^{2}}{M_{Pl}^{2}} \right).
\end{eqnarray}
These are equivalent to the results obtained from Palatini formalism \cite{bauer}.

The number of $e$-foldings is given by
\begin{eqnarray}
N=\frac{1}{M^{2}_{Pl}}\int_{h_{\text{end}}}^{h_{\text{start}}}
\frac{U}{d U/d h}\left(\frac{d \chi}{d h} \right)^{2} \nonumber
\simeq\frac{1}{32\xi}\exp \left(\frac{\xi h_{\text{start}}^{2}}{M^{2}_{Pl}} \right). \label{efoldings}
\end{eqnarray}
Here the final field $h_{\text{end}}$ corresponds to the end of inflation where the slow roll conditions break down, or $\epsilon \simeq 1$, and the initial field $h_{\text{start}}$ is determined from the number of $e$-foldings $N$.

For the number of $e$-folds $N=50-70$, and at first order, the spectral index $n_{s}$ is in the range
\begin{eqnarray}
0.960 \leq n_{s} \leq 0.970,
\end{eqnarray}
which is in agreement with the recent Planck results.

Planck data constraint on the power spectrum of the primordial perturbations generated during inflation is given by \cite{planck}
\begin{eqnarray}
\frac{H^{2}}{8\pi^{2} \epsilon M^{2}_{Pl}} \simeq 2.4 \times 10^{-9},
\end{eqnarray}
which leads to
\begin{eqnarray}
\frac{\lambda}{\xi}   \simeq 2.66 \times 10^{-11}.
\end{eqnarray}
Then, the SM quartic coupling $\lambda \simeq 0.13$ implies
\begin{eqnarray}
\xi \simeq 4.8 \times 10^{9}.
\end{eqnarray}
The affine nonminimal coupling is then larger than its value in metric gravity. This leads to an extremely small tensor-to-scalar ratio
\begin{eqnarray}
r=16\epsilon \simeq \mathcal{O}\left(10^{-13} \right).
\end{eqnarray}
As we see, although the predicted spectral index agrees with the measured value, the tensor contribution is tiny and negligible. This is because of the flatness of the potential (\ref{field and potential transformation}). Recent observations suggest a very small upper bound for tensor perturbations, the tensor-to-scalar ratio is of the order $r < 0.08 $. Future observations are expected to provide us with a precise bounds, since then, one may decide whether Higgs affine inflation could be considered as a good model for the early universe. In Table~\ref{tab:2} we summarize the results obtained here and compare them with Higgs inflation in metric gravity.

\begin{table}[h]
\centering
\begin{tabular}{|c|c|cc|}
\hline
Parameters &Higgs Inflation (metric gravity) &Higgs Affine Inflation&\\
\hline
\, & \, & \, &\\
$\xi$ & $10^{4}$ & $10^{9}$& \\
\, & \, & \, &\\
$n_{s}$& $0.97$ & $0.97$ &\\
\, & \, & \, &\\
$r$ & $0.0032$ & $\mathcal{O}(10^{-13})$ &\\
\hline
\end{tabular}
\caption{\label{tab:2} Predicted parameters based on SM Higgs inflation in both metric and affine gravity. Higgs affine inflation requires a strong Higgs-curvature coupling but a negligible tensor-to-scalar ratio. }
\end{table}

\newpage
\subsection{Starobinsky affine inflation}
\label{starobinski inflation}
In metric theories of gravity, the $R^{2}$ inflationary model is based on the following action \cite{starobinsky}
\begin{eqnarray}
\label{starobinsky metric action}
S=\frac{1}{2}\int d^{4}x \sqrt{\left |\right| g \left |\right|}
 \left[M^{2}_{Pl} R(g)+\frac{R^{2}(g)}{6M^{2}}  \right],
\end{eqnarray}
where $M$ is of mass dimension.

The usual problem with this type of theories is that they would invoke higher order derivatives (up to forth order here). In its original form, the theory (\ref{starobinsky metric action}) describes the propagation of spin-2 state, however, it can be shown that this theory may be derived by integrating out a scalar degree of freedom. It is convenient then to introduce a scalar field $\phi$ and perform a conformal transformation where the field $\phi$ is minimally coupled to gravity in Einstein frame. This transformation is given by 
\begin{eqnarray}
g_{\mu\nu} \rightarrow e^{-\sqrt{\frac{2}{3}}\,\frac{\phi}{M_{Pl}}} \, g_{\mu\nu}.
\end{eqnarray}
This makes the Starobinski model (\ref{starobinsky metric action}) equivalent to a theory of a scalar field coupled to gravity in Einstein frame as
\begin{eqnarray}
S=\int d^{4}x \sqrt{\left |\right| g \left |\right|}
\left[\frac{M^{2}_{Pl}}{2}R(g)-(\nabla \phi)^{2}-\frac{3}{4}M^{4}_{Pl}M^{2}
\left(1-e^{-\sqrt{\frac{2}{3}}\,\frac{\phi}{M_{Pl}}} \right)^{2} \right].
\end{eqnarray}
It is the flat potential which appears in the last term that drives inflation. The model predicts a scalar tilt and a small tensor-to-scalar ratio which are consistent with Planck constraints \cite{predictions of starobinsky}
\begin{eqnarray}
\label{starobinski tilt and ratio}
n_{s}\simeq 1-\frac{2}{N} \, \quad \quad \text{and}\quad \quad r\simeq \frac{12}{N^{2}},
\end{eqnarray}
where the parameter $M \simeq 10^{-5}$ is fixed by the normalization of the CMB anisotropies.

As we have seen so far, affine gravity generates a unique metric tensor where the gravitational equations are equivalent to Einstein equations for a minimal coupled field ($\xi=0$). A scalar field $\phi$ coupled minimally to affine gravity via the action (\ref{transformed affine action}) leads to the same results (\ref{starobinski tilt and ratio}) if it is associated with a potential of the form
\begin{eqnarray}
V\left(\phi\right)=
\frac{3}{4}M^{4}_{Pl}M^{2}
\left(1-e^{-\sqrt{\frac{2}{3}}\,\frac{\phi}{M_{Pl}}} \right)^{2}.
\end{eqnarray}
This is clearly a consequence of the equivalence of Einstein's general relativity and affine gravity in case of minimal coupling. In the general case ($\xi \neq 0$), the two theories are no longer equivalent and then we expect different predictions.

\subsection{Affine $\alpha$-attractors}
\label{subsec: alpha-attractor}
Most of the successful inflationary models are based on flat potentials. Although, they realize the slow-roll inflation, steeper potentials like the quadratic potential provide a large tensor-to-scalar ratios which are not in agreement with the observed results. This led people in the recent few years to illuminating the old chaotic models of inflation by modifying the dynamics of the inflaton through its kinetic couplings. These $\alpha$-attractor models which are motivated from supergravity become indeed of great interest since they provide excellent fits to observation \cite{alpha-attractors}. These models are implemented in their standard form in the context of metric gravity, and it may be worth shedding light on their realization in the context of affine gravity.

Up to now, our scalar fields $\phi$ (single and mutifields) are put in our setup in their canonical kinetic terms in both cases, minimal and nonminimal couplings. Here, and for a general case, the canonical form may be broken, and one may write a general affine action of the form
\begin{eqnarray}
\label{non canonical coupling}
S[\Gamma,\phi]=
\int d^{4}x \frac{\sqrt{||M^{2}_{Pl} \mathcal{F}(\phi)R_{\mu\nu}(\Gamma)-G(\phi)\nabla_{\mu}\phi\nabla_{\nu}\phi||}}{V(\phi)},
\end{eqnarray}     
where we have introduced the nonzero function $G(\phi)$ as the breaking source of the canonical form. There are two important cases in this setup:
\begin{itemize}
\item \textit{Minimal but non-canonical}: 

The first case is when the function $\mathcal{F}(\phi)\rightarrow 1$. It is this case that realizes a simple $\alpha$-attractor model, in fact, the field $\phi$ is now rescaled as
\begin{eqnarray}
d\phi \rightarrow \frac{d\varphi}{\sqrt{G(\phi)}},
\end{eqnarray} 
leading to a canonical field $\varphi$ coupled minimally to affine gravity via 
\begin{eqnarray}
\label{canonical coupling}
S[\Gamma,\varphi]=
\int d^{4}x \frac{\sqrt{||M^{2}_{Pl}R_{\mu\nu}(\Gamma)-\nabla_{\mu}\varphi\nabla_{\nu}\varphi||}}{V[\phi(\varphi)]}.
\end{eqnarray} 
This theory is equivalent to the $\alpha$-attractor model studied in metric theory, if one similarly takes the following function \cite{alpha-attractors}
\begin{eqnarray}
\label{function G}
G(\phi)=\frac{1}{\left(1-\frac{\phi^{2}}{6\alpha} \right)^{2}},
\end{eqnarray}
where $\alpha$ is a constant which is taken small for observational reason.

In this case, and for a simple quadratic potential, the scalar index and tensor-to-scalar ratio are given in terms of $\alpha$ and the number of $e$-foldings by
\begin{eqnarray}
n_{s}\simeq 1-\frac{2}{N}, \, \, \quad r \simeq \frac{12\alpha}{N^{2}}.
\end{eqnarray}
What is interesting in these models is that they provide a plateau potential for any non-singular potential $V(\phi)$
\begin{eqnarray}
V(\phi) \rightarrow 
V=V\left[\tanh\left(\frac{\varphi}{\sqrt{6\alpha}} \right)  \right],
\end{eqnarray}
when switching to the canonical field $\varphi$.

This example may look trivial since it is based on the function (\ref{function G}) which has been proposed in metric theory \cite{alpha-attractors}. However, the gravity theory is different and this example arises only as a particular case and it can be considered as a different realization of $\alpha$ attractor models.  
\item 
\textit{Nonminimal and non-canonical}:

This case is general and it describes a non-canonical field coupled nonminimally to affine gravity. Here switching to a canonical field $\varphi$ must be followed by the transformation to minimal coupling. As we have seen throughout the review, this can be done easily in affine gravity using a field redefinition. However, there is another simple and interesting case where the functions coincide, $\mathcal{F(\phi)}=G(\phi)$. In this case, not the inflaton but only the potential which is modified
\begin{eqnarray}
V(\phi) \rightarrow V(\phi)=\frac{V(\phi)}{G^{2}(\phi)}.
\end{eqnarray}
This example shows that the potential could be made flat in terms of the original field $\phi$ which has taken a canonical form after the transformation. It may be difficult (if not impossible) to realize this property in metric gravity.
\end{itemize}

\section{Concluding remarks and new insights}
\label{sec:conclusion}

Conformal frames in gravitational theories are traced back to Jordan who proposed an extended theory of gravity, named after that, scalar-tensor theory of gravity \cite{jordan}. It was shown afterwards by Brans and Dicke that these theories are equivalent to Einstein's general relativity with a scalar field when some rescalings are applied on the metric tensor and the old scalar field \cite{brans}. Since then people realized that the transition made between the two theories has led to the existence of two possible distinct frames, Jordan and Einstein frames. 

The goal of the present work is not to solve the problem of frames and decide whether Jordan or Einstein frame is physical, indeed the problem of conformal frames arises in metric theory of gravity like GR and it may be restricted to it. However, the present work can be considered as a new setup towards avoiding the use of conformal frames themselves. Affine gravity as we have seen throughout this review provides us with an origin to the metric tensor. This tensor is unique in a sense that it arises from both minimal and nonminimal couplings of scalar fields to affine connection and its curvature. A unique description of the gravitational sector prevents the use of conformal frames. While transition between different couplings is performed via rescaling of the scalar fields, the geometric quantities like the Hubble parameter, remain unchanged, this leads to the invariance of the power spectra produced by different inflatons.

Purely affine gravity is not a new theory, it goes back to previous classic works of Einstein, Eddington and Schr{\"{o}}dinger as an attempts to a unified picture of gravity and electrodynamics \cite{eddington}. The failure of this purpose of unification has led people to abandoning the affine approach by considering it as a pure mathematical construction that lacks physical interpretations. Other affine approach to gravity has been proposed later as a different formulation of general relativity where the metric tensor appears as a momentum canonical conjugate to the affine connection, and the derived field equations are equivalent to those of GR with scalar and possibly gauge fields \cite{kijowski1,kijowski2}. In the recent few years, attempts have been made to consider general and different approaches to pure affine gravity, in vacuum and in the presence of matter and even in higher dimensions \cite{demir-eddington,kemal,azri-eddington,azri-separate,liebscher,oscar,poplawski}. 

As we have shown in this work, the affine dynamics in the presence of scalar fields may naturally be applied to physical phenomena. This is clearly seen in the case of cosmic inflation where deviations from metric theory are remarkable \cite{affine inflation,induced affine inflation}. Furthermore, we have argued and shown that metric gravity itself may have arisen from affine spacetime that incorporates only scalar fields with a nonzero potential. In our opinion, this feature could be a convincing reason in order to pay much more attention to affine gravity. 

Affine gravity provides a possible viable description of the early universe since it accommodates scalar fields (inflaton) and imposes a nonzero potential energy. We believe that the SM matter fields which we have not considered here may be incorporated into the theory in a satisfactory manner providing a complete and unified picture of the SM in affine spacetime. Another possible and new insight is that the SM matter fields, although difficult to be incorporated directly in the setup, they may be generated dynamically at the end of inflation where the inflaton energy is converted to SM particles and the universe becomes radiation dominated. This \enquote{speculative} mechanism needs to be studied as a reheating process in the context of affine gravity. Last but not least, the quantum correction to the affine actions is not trivial, these actions are not polynomials in the fields and one might go beyond the standard techniques when performing the covariant quantization. However, it may be possible to convert these actions into polynomials that lead to the same equations of motion, but in this case one may lose the aim of affine gravity by proposing different forms of the action \cite{quantum eddington}.   

\section*{Acknowledgements}
The work was supported in part by the T{\"U}B{\.I}TAK Grant 115F212. The author thanks Durmu{\c{s}} Ali Demir for suggesting this short review and for helpful discussions during its preparation. He also thanks David Kaiser for pointing out useful references on inflation in the context of scalar-tensor theories, and Andrei Linde for discussing inflation. Part of this work was completed during the Nordita program; \textit{Advances in Theoretical Cosmology in Light of Data}, and the author acknowledges the warm hospitality there.  

\appendix

\section{CONFORMAL TRANSFORMATION}
\label{appendix1}
Without any abstract mathematical definition, a conformal transformation in the spacetime manifold is the mapping that allows the transition between two metric tensors $g_{\mu\nu}$ and $\tilde{g}_{\mu\nu}$ via the following relation
\begin{eqnarray}
\tilde{g}_{\mu\nu}=\mathcal{F} g_{\mu\nu},
\end{eqnarray}
where $\mathcal{F}$ is a function of spacetime coordinates.

Since this transformation is not coordinate transformation, then the differentials $dx^{\mu}$ are not subjected to it. In this case, one may write this transformation in terms of the line elements as
\begin{eqnarray}
d\tilde{s}^{2}=\mathcal{F} ds^{2}.
\end{eqnarray} 
To keep the same sign for the line element, we usually take a positive function $\mathcal{F}=\Omega^{2}$. 

A conformal transformation can be represented then as an isotropic expansion or contraction \cite{maeda book,wald}.
 
The inverse transformation is easily written as
\begin{eqnarray}
\tilde{g}^{\mu\nu}=\mathcal{F}^{-1} g^{\mu\nu},
\end{eqnarray}
and the scalar density which defines the volume element transforms as
\begin{eqnarray}
\sqrt{||\tilde{g}||}=\mathcal{F}^{2}\sqrt{||g||}.
\end{eqnarray}
The metric transformations impose the following transformation on the Levi-civita connection
\begin{eqnarray}
\tilde{\Gamma}^{\alpha}_{\mu\nu}=
\Gamma^{\alpha}_{\mu\nu}
+\frac{1}{2}\mathcal{F}^{-1}
\Big( \delta^{\alpha}_{\mu} \nabla_{\nu} \mathcal{F} + \delta^{\alpha}_{\nu} \nabla_{\mu} \mathcal{F}-g_{\mu\nu} \nabla^{\alpha} \mathcal{F}
\Big).
\end{eqnarray}
To obtain the Riemann tensor, one may first write the derivative of this connection which transforms as
\begin{eqnarray}
\partial_{\beta} \tilde{\Gamma}^{\alpha}_{\mu\nu}=&
\partial_{\beta} \Gamma^{\alpha}_{\mu\nu}
-\frac{1}{2}\mathcal{F}^{-2} \nabla_{\beta}{\mathcal{F}^{2}}
\Big( \delta^{\alpha}_{\mu}\nabla_{\nu}\mathcal{F}
+\delta^{\alpha}_{\nu}\nabla_{\mu}\mathcal{F}
-g_{\mu\nu}\nabla^{\alpha}\mathcal{F}
\Big) \nonumber \\
& +\frac{1}{2}\mathcal{F}^{-1}
\Big( \delta^{\alpha}_{\mu} \nabla_{\nu}\nabla_{\beta} \mathcal{F}
+\delta^{\alpha}_{\nu} \nabla_{\mu}\nabla_{\beta} \mathcal{F}
-g_{\mu\nu} \partial_{\beta} g^{\alpha\gamma}\nabla_{\gamma} \mathcal{F}
\nonumber \\
&-\partial_{\beta}g_{\mu\nu} \nabla^{\alpha}\mathcal{F}
-g_{\mu\nu} \nabla^{\alpha}\nabla_{\beta}\mathcal{F}
\Big),
\end{eqnarray}
and finally write the transformed Riemann tensor
\begin{eqnarray}
\tilde{R}^{\alpha}_{\,\,\beta\gamma\delta}(\tilde{\Gamma})=&
R^{\alpha}_{\,\,\beta\gamma\delta}(\Gamma)
-\frac{1}{2} \mathcal{F}^{-1}
\Big(g_{\beta\gamma} \nabla^{\alpha} \nabla_{\delta}\mathcal{F}
-g_{\beta\delta} \nabla^{\alpha} \nabla_{\gamma}\mathcal{F}
+\delta^{\alpha}_{\delta} \nabla_{\beta}\nabla_{\gamma}\mathcal{F}
-\delta^{\alpha}_{\gamma} \nabla_{\beta}\nabla_{\delta}\mathcal{F}
\Big) \nonumber \\
&+\frac{1}{4}\mathcal{F}^{-1}
\Big( 3g_{\beta\delta}
\nabla_{\gamma}\mathcal{F} \nabla^{\alpha}\mathcal{F}
-3g_{\beta\gamma} \nabla_{\delta}\mathcal{F} \nabla^{\alpha}\mathcal{F}
+3\delta^{\alpha}_{\gamma} \nabla_{\beta}\mathcal{F} \nabla^{\delta}\mathcal{F}
-3\delta^{\alpha}_{\delta} \nabla_{\beta}\mathcal{F} \nabla^{\gamma}\mathcal{F} \nonumber \\
&+g_{\beta\gamma}\delta^{\alpha}_{\delta} \nabla_{\nu}\mathcal{F} \nabla^{\nu}\mathcal{F}
-g_{\beta\gamma}\delta^{\alpha}_{\gamma} \nabla_{\nu}\mathcal{F} \nabla^{\nu}\mathcal{F}
\Big).
\end{eqnarray}
The easiest work now is to contract the Riemann tensor and get both Ricci tensor and Ricci scalar respectively
\begin{eqnarray}
\tilde{R}_{\alpha\beta}(\tilde{\Gamma})=
R_{\alpha\beta}(\Gamma)
+\frac{3}{2} \mathcal{F}^{-2}\nabla_{\alpha}\mathcal{F}\nabla_{\beta}\mathcal{F}
-\mathcal{F}^{-1}\nabla_{\alpha}\nabla_{\beta}\mathcal{F}
-\frac{1}{2}g_{\alpha\beta} \mathcal{F}^{-1}\Box \mathcal{F}.
\end{eqnarray}
\begin{eqnarray}
\label{transformed ricci scalar}
\tilde{R}(\tilde{\Gamma})&&=\tilde{g}^{\alpha\beta}\tilde{R}_{\alpha\beta}(\tilde{\Gamma}) \nonumber \\
&&=\mathcal{F}^{-1}R(\Gamma) +\frac{3}{2}\mathcal{F}^{-3} \nabla^{\nu}\mathcal{F}
\nabla_{\nu}\mathcal{F}- 3\mathcal{F}^{-2}\Box \mathcal{F}.
\end{eqnarray}
If the function $\mathcal{F}$ is a function of a physical field $\phi$, the second term in the right hand side of (\ref{transformed ricci scalar}) would be proportional to the kinetic term of the field. 

Equation (\ref{transformed ricci scalar}) is the relation which is used in the transition between Jordan and Einstein frames. It is clear that Einstein's equations are not invariant under conformal transformation.

\section{AFFINE DYNAMICS }
\subsection{Invariant actions and variation}
\label{appendix2}
Herein, the local properties of spacetime are completely specified by the affine connection $\Gamma_{\alpha\beta}^{\lambda}$ and the associated curvature. What we have called metric tensor in the previous appendix does not make any sense here, this tensor may be only assumed and it adds an extra geometric concept (metric structure) which will not be supposed here.  

In general, this connection is asymmetric and then the torsion tensor plays an important role. However, as we have done thorough out this paper, we will choose a symmetric affine connection and use the symmetric part of the curvature.

The Riemann tensor is defined as
\begin{eqnarray}
R^{\lambda}_{\,\,\alpha\mu\beta}(\Gamma)
=\partial_{\mu} \Gamma_{\alpha\beta}^{\lambda}
-\partial_{\beta} \Gamma_{\alpha\mu}^{\lambda}
+\Gamma_{\sigma\mu}^{\lambda}\Gamma_{\alpha\beta}^{\sigma}
-\Gamma_{\sigma\beta}^{\lambda}\Gamma_{\alpha\mu}^{\sigma}.
\end{eqnarray}
This leads to the Ricci tensor when summing the indices $\lambda$ and $\mu$
\begin{eqnarray}
R_{\alpha\beta}(\Gamma)&&= R^{\lambda}_{\,\,\alpha\lambda\beta}(\Gamma) \nonumber
\\
&&=\partial_{\lambda} \Gamma_{\alpha\beta}^{\lambda}
-\partial_{\beta} \Gamma_{\alpha\lambda}^{\lambda}
+\Gamma_{\sigma\lambda}^{\lambda}\Gamma_{\alpha\beta}^{\sigma}
-\Gamma_{\sigma\beta}^{\lambda}\Gamma_{\alpha\lambda}^{\sigma}
\end{eqnarray}
In the affine calculus of variation, the invariant action will be varied with respect to the affine connection. For the Ricci tensor, this reads
\begin{eqnarray}
\delta R_{\alpha\beta}=
\partial_{\lambda} (\delta \Gamma_{\alpha\beta}^{\lambda})
-\partial_{\beta} (\delta \Gamma_{\alpha\lambda}^{\lambda})
+\delta\Gamma_{\sigma\lambda}^{\lambda}\Gamma_{\alpha\beta}^{\sigma}
+\Gamma_{\sigma\lambda}^{\lambda}\delta\Gamma_{\alpha\beta}^{\sigma}
-\delta \Gamma_{\sigma\beta}^{\lambda}\Gamma_{\alpha\lambda}^{\sigma}
-\Gamma_{\sigma\beta}^{\lambda}\delta \Gamma_{\alpha\lambda}^{\sigma}.
\end{eqnarray}
Unlike $\Gamma_{\alpha\beta}^{\lambda}$, the coefficients $\delta \Gamma_{\alpha\beta}^{\lambda}$ are not components of a connection but rather, they define a tensor. Thus, one may apply the covariant derivative on this tensor and easily show the important property 
\begin{eqnarray}
\label{ricci variation}
\delta R_{\alpha\beta}=
\nabla_{\lambda} (\delta \Gamma_{\alpha\beta}^{\lambda})
-\nabla_{\beta} (\delta \Gamma_{\alpha\lambda}^{\lambda}).
\end{eqnarray}
Since our spacetime is not endowed with a metric tensor, this makes defining the invariant actions seems problematic. However, all what we need is an invariant measure (volume element) which is generally defined by the square-root of the determinant of a rank-two tensor. Our affine spacetime contains curvature and derivatives of matter field, these quantities may be used to define a simple rank-two tensor as
\begin{eqnarray}
\label{tensor K(phi)}
K_{\alpha\beta}(\Gamma,\phi)=
\mathcal{F}(\phi)R_{\alpha\beta}(\Gamma)-\nabla_{\alpha}\phi\nabla_{\beta}\phi.
\end{eqnarray}
The presence of the potential energy $V(\phi)$ completes our setup by proposing the invariant action
\begin{eqnarray}
\label{general affine action}
S[\Gamma,\phi]=
\int d^{4}x \frac{\sqrt{\left|\left| K(\Gamma,\phi)\right|\right|}}{V(\phi)}.
\end{eqnarray}
Now, using the property (\ref{ricci variation}), the variation of this action ($\delta S=0$) implies
\begin{eqnarray}
\int d^{4}x
\frac{\sqrt{\left|\left| K(\Gamma,\phi)\right|\right|}}{V(\phi)}
\mathcal{F}(\phi)
(K^{-1})^{\alpha\beta}
\left(\nabla_{\lambda} (\delta \Gamma_{\alpha\beta}^{\lambda})
-\nabla_{\beta} (\delta \Gamma_{\alpha\lambda}^{\lambda})  \right)=0.
\end{eqnarray}
By integrating by parts and getting rid of the surface terms, we obtain
\begin{eqnarray}
\int d^{4}x \Bigg[ \nabla_{\nu}\left(\frac{\sqrt{\left|\left| K(\Gamma,\phi)\right|\right|}}{V(\phi)}
\mathcal{F}(\phi)
(K^{-1})^{\mu\nu} \delta_{\lambda}^{\kappa}\delta_{\mu}^{\sigma} \right) -\nabla_{\lambda}\left(\frac{\sqrt{\left|\left| K(\Gamma,\phi)\right|\right|}}{V(\phi)}
\mathcal{F}(\phi)
(K^{-1})^{\mu\nu} \delta_{\mu}^{\kappa}\delta_{\nu}^{\sigma} \right)
\Bigg] \delta \Gamma_{\kappa\sigma}^{\lambda}=0 \nonumber 
\end{eqnarray}
\begin{eqnarray}
\,
\end{eqnarray}
This leads to the dynamical equation
\begin{eqnarray}
\nabla_{\nu}\left(\frac{\sqrt{\left|\left| K(\Gamma,\phi)\right|\right|}}{V(\phi)}
\mathcal{F}(\phi)
(K^{-1})^{\sigma\nu} \right)\delta_{\lambda}^{\kappa}
-\nabla_{\lambda}\left(\frac{\sqrt{\left|\left| K(\Gamma,\phi)\right|\right|}}{V(\phi)}
\mathcal{F}(\phi)
(K^{-1})^{\kappa\sigma}\right)=0,
\end{eqnarray}
which is equivalent to
\begin{eqnarray}
\nabla_{\alpha}\left(\mathcal{F}(\phi) \frac{\sqrt{\left|\left| K(\Gamma,\phi)\right|\right|}}{V(\phi)}
(K^{-1})^{\mu\nu} \right)=0.
\end{eqnarray}
This equation is the basis of all the affine gravity models presented in this paper.

The metric tensor then arises as a solution to this equation
\begin{eqnarray}
\label{general metric}
\sqrt{\left|\left| g \right| \right|}(g^{-1})^{\mu\nu}
=\mathcal{F}(\phi) \frac{\sqrt{\left|\left| K(\Gamma,\phi)\right|\right|}}{V(\phi)}
(K^{-1})^{\mu\nu}.
\end{eqnarray}
When written in a tensor form, the last equality leads to the gravitational equations.

Now, variation of action (\ref{general affine action}) with respect to the field $\phi$ implies
\begin{eqnarray}
\int d^{4}x
\Biggl[
\frac{1}{2}\frac{\sqrt{\left|\left| K(\Gamma,\phi)\right|\right|}}{V(\phi)}
(K^{-1})^{\alpha\beta} \left[\mathcal{F}^{\prime}(\phi)\delta \phi R_{\alpha\beta}
-\nabla_{\alpha}\phi \nabla_{\beta}(\delta \phi)-\nabla_{\alpha}(\delta \phi) \nabla_{\beta}\phi  \right] \nonumber \\
-\frac{\sqrt{\left|\left| K(\Gamma,\phi)\right|\right|}}{V^{2}(\phi)}V^{\prime}(\phi)\delta \phi
\Biggr]=0
\end{eqnarray}
Again, by integrating by parts the terms containing the derivatives of the field, and getting rid of the surface terms, we obtain
\begin{eqnarray}
\int d^{4}x \Biggl[
\partial_{\beta}\left(\frac{\sqrt{\left|\left| K(\Gamma,\phi)\right|\right|}}{V(\phi)}(K^{-1})^{\alpha\beta} \partial_{\alpha}\phi \right)
-\frac{\sqrt{\left|\left| K(\Gamma,\phi)\right|\right|}}{V^{2}(\phi)}V^{\prime}(\phi) \nonumber \\
+\frac{1}{2}\mathcal{F}^{\prime}(\phi) \frac{\sqrt{\left|\left| K(\Gamma,\phi)\right|\right|}}{V(\phi)}
(K^{-1})^{\alpha\beta} R_{\alpha\beta}
\Biggr]
\delta \phi=0,
\end{eqnarray}
which gives the equation
\begin{eqnarray}
\partial_{\beta}\left(\frac{\sqrt{\left|\left| K(\Gamma,\phi)\right|\right|}}{V(\phi)}(K^{-1})^{\alpha\beta} \partial_{\alpha}\phi \right)
-\frac{\sqrt{\left|\left| K(\Gamma,\phi)\right|\right|}}{V^{2}(\phi)}V^{\prime}(\phi)
+\frac{1}{2}\mathcal{F}^{\prime}(\phi) \frac{\sqrt{\left|\left| K(\Gamma,\phi)\right|\right|}}{V(\phi)}
(K^{-1})^{\alpha\beta} R_{\alpha\beta}=0 \nonumber
\end{eqnarray}
\begin{eqnarray}
\,
\end{eqnarray}
In terms of the metric tensor (\ref{general metric}), and after some simplifications, the last equations are brought to a standard form as
\begin{eqnarray}
\Box \phi -V^{\prime}(\phi)+\frac{1}{2}\mathcal{F}^{\prime}(\phi)R(g)+\Psi(\phi)=0,
\end{eqnarray}
where
\begin{eqnarray}
\Psi(\phi)=\left(1-\mathcal{F}^{-1} \right)V^{\prime}(\phi)-\mathcal{F}^{-1}\mathcal{F}^{\prime}(\nabla \phi)^{2}.
\end{eqnarray}

\subsection{Field redefinition and uniqueness of the metric}

For simplicity, we may take $M^{2}_{Pl}=1$. The nonminimally coupled field $\phi$ is transformed to minimally coupled field $\tilde{\phi}$ using the following redefinition
\begin{eqnarray}
\label{transformation phi to tildephi}
d\tilde{\phi}=\frac{d\phi}{\sqrt{\mathcal{F}}(\phi)},\,\,\,
\tilde{V}(\tilde{\phi})=\frac{V(\phi)}{\mathcal{F}^{2}(\phi)}.
\end{eqnarray}
Under this transformation, the invariant action (\ref{general affine action}) reads
\begin{eqnarray}
\label{transformed general affine action}
S[\Gamma,\phi]
=\int d^{4}x \frac{\sqrt{| | \tilde{K}(\Gamma,\tilde{\phi}) | |}}
{\tilde{V}(\tilde{\phi})},
\end{eqnarray}
where
\begin{eqnarray}
\label{K(tildephi)}
\tilde{K}_{\mu\nu}(\Gamma,\tilde{\phi})= R_{\mu\nu}(\Gamma)
-\nabla_{\mu}\tilde{\phi}\nabla_{\nu}\tilde{\phi}.
\end{eqnarray}
Following the same steps in varying action (\ref{general affine action}), in this time variation of action (\ref{transformed general affine action}) with respect to $\Gamma$ leads to the dynamical equation
\begin{eqnarray}
\nabla_{\alpha}\left(\frac{\sqrt{||\tilde{ K}(\Gamma,\tilde{\phi})||}}{\tilde{V}(\tilde{\phi})}
(\tilde{K}^{-1})^{\mu\nu} \right)=0.
\end{eqnarray}
Again a metric tensor $\tilde{g}_{\mu\nu}$ arises as a solution of the last equation
\begin{eqnarray}
\label{tilde general metric}
\sqrt{|| \tilde{g} ||}(\tilde{g}^{-1})^{\mu\nu}
=\frac{\sqrt{|| \tilde{K}(\Gamma,\tilde{\phi})||}}{\tilde{V}(\tilde{\phi})}
(\tilde{K}^{-1})^{\mu\nu}.
\end{eqnarray}
If we use (\ref{transformation phi to tildephi}) again, and relation (\ref{general metric}) we get
\begin{eqnarray}
\label{metric equality}
\sqrt{-\tilde{g}}\tilde{g}^{\mu\nu}&&=
\mathcal{F}(\phi)\frac{\sqrt{\left| \right| K_{\mu\nu}\left(\phi \right) \left| \right| }}{V\left(\phi\right)} \left( K^{-1} \right)^{\mu\nu} \nonumber \\
&&\equiv \sqrt{-g}g^{\mu\nu},
\end{eqnarray}
Which shows that the generated metric tensor is unique.

\end{document}